\documentclass[10pt,journal,compsoc]{IEEEtran}
\IEEEoverridecommandlockouts
\usepackage{cite}
\usepackage{amsmath,amssymb,amsfonts}
\usepackage{algorithmic}
\usepackage{graphicx}
\usepackage{textcomp}
\usepackage{bm}
\usepackage{makecell}
\usepackage{hhline}
\usepackage{caption}
\usepackage[table]{xcolor} 
\usepackage{booktabs} 
\usepackage{hyperref}
\usepackage{balance} 
 \usepackage{soul} 
 \usepackage{xcolor}
 
\definecolor{gray}{rgb}{0.35, 0.35, 0.35}

\hypersetup{
    colorlinks=true,   
    linkcolor=blue,    
    citecolor=blue,    
    urlcolor=blue,      
}

\def\BibTeX{{\rm B\kern-.05em{\sc i\kern-.025em b}\kern-.08em
		T\kern-.1667em\lower.7ex\hbox{E}\kern-.125emX}}

\usepackage{graphicx}
\usepackage{booktabs}
\usepackage{multirow}
\usepackage{subcaption}
\usepackage{tabularx}
\usepackage[ruled,linesnumbered]{algorithm2e} 
\usepackage{pifont}

\def\bf{\boldsymbol{f}}

\definecolor{frenchblue}{rgb}{0.0, 0.45, 0.73}

\usepackage{color}

\begin{document}
	
	\title{
	\textcolor{black}{From Signals to Evidence: Interpretable Underwater Acoustic Target Sensing\\ with Large Language Models \\— Part I: Framework and Recognition}
	}

    \author{
    \thanks{\rm \textcolor{black}{The code is released at \url{https://github.com/xilyao/CFD-DC}.}}
    }
	

		\author{Yuge Zhang, Jie Chen, Zhengqiao Zhao, Xiling Yao, Shumin Liu\\Lingji Xu, Jingdong Chen, Yixin Yang
		\IEEEcompsocitemizethanks{
\IEEEcompsocthanksitem Corresponding author: Jie Chen.
    \IEEEcompsocthanksitem   

     Y. Zhang, J. Chen, Z. Zhao, X. Yao, S. Liu, J.-D. Chen are with School of Artificial Intelligence, Northwestern Polytechnical University, Xi’an, Shaanxi 710129, China (emails: zhangyuge@mail.nwpu.edu.cn, dr.jie.chen@ieee.org, zhengqiao.zhao@nwpu.edu.cn, xilingyao@mail.nwpu.edu.cn, liu\_shumin@nwpu.edu.cn, jchen@nwpu.edu.cn).

     L. Xu is with School of Ocean Engineering and Technology, Sun Yat-sen University, Zhuhai, 519000, Guangdong, China (xulj26@mail.sysu.edu.cn).

    Y. Yang is with School of Marine Science and Technology, Northwestern Polytechnical University, Xi’an, Shaanxi 710129, China, and also with School of Electronic and Optical Engineering, Nanjing University of Science and Technology, Nanjing, Jiangsu 210094, China (email: yxyang@nwpu.edu.cn).
    }
    }

	\IEEEtitleabstractindextext{
	\begin{abstract}
		Underwater acoustic target detection and recognition are essential to marine surveillance and intelligent ocean sensing. However, existing deep learning systems typically provide binary decisions, categorical labels, or confidence scores, offering limited insight into the physical evidence behind their outputs. Although large language models (LLMs) offer new opportunities for interpretable acoustic understanding, generic audio encoders often fail to capture underwater-specific physical characteristics, while the scarcity of physically annotated data remains a major obstacle. In this paper, the first of a two-part series, we propose an Underwater acoustic Physics-aware Description (UPD) framework that integrates physics-aware acoustic encoding with a pretrained LLM through a unified feature–evidence–decision chain. Its open, extensible input layer accommodates heterogeneous information, including raw waveforms, environmental and geographic data, and textual physical priors, enabling detection and recognition to be formulated within a common framework. In In Part I, we instantiate this framework for target recognition through an Underwater acoustic Physics-aware Transformer Encoder(UPTE), which tokenizes complementary acoustic representations and learns task-disentangled physical representations through four physics-oriented learning tasks. To support this process, we develop a physics-parameterized data engine that provides traceable supervision at two levels of granularity and introduce an attribute-level evaluation protocol to assess the physical fidelity of generated descriptions. Experiments on simulated recordings of four vessel types demonstrate accurate recognition supported by physically grounded attributes. The framework also handles unknown-class samples by identifying their physical characteristics without assigning them to predefined vessel categories.

	\end{abstract}
	
	}
	\maketitle
	\IEEEdisplaynontitleabstractindextext
	
	\IEEEpeerreviewmaketitle
	
	\section{Introduction}
	\label{sec:intro}
	
Passive underwater acoustic sensing forms a cornerstone of marine surveillance, maritime traffic safety, and intelligent ocean observation~\cite{hummel2024survey,helal2024gliders}. Its processing chain consists of two main stages: detection and recognition. Detection determines whether a target is present in the received acoustic field, often under low signal-to-noise ratio (SNR) conditions. Recognition then identifies the target by analyzing faint, structured acoustic signatures in ship-radiated noise, including tonal lines, envelope modulations, and machinery-related spectral components~\cite{guo2024modulation,hanson2008demon}. The two stages are intimately coupled: detection acts as the front-end gatekeeper, close to the raw acoustic field where signatures are weakest, while recognition operates downstream on detected contacts. These signatures are severely attenuated and distorted during propagation, while different vessel categories can exhibit similar spectral and modulation patterns. These factors make reliable detection and robust recognition long-standing challenges in underwater acoustics~\cite{jensen2011computational,hummel2024survey}. Interpretability is becoming increasingly important throughout this processing chain. Operators and autonomous systems need to understand not only the system’s decisions but also the physical evidence underlying them. However, both detection and recognition remain largely reliant on methods that offer limited insight into their decision-making processes.

\begin{figure*}[!t]
	\centering
	\includegraphics[width=1\textwidth]{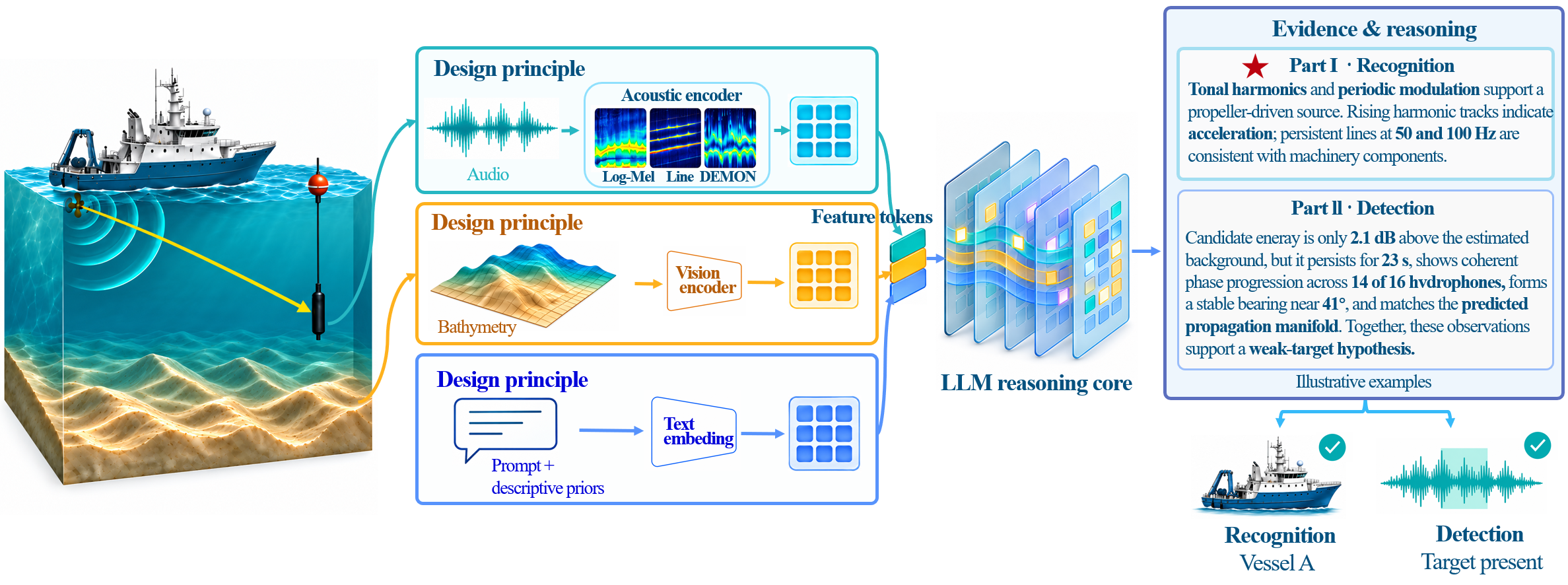}
	\caption{Conceptual illustration of the unified UPD framework for underwater acoustic detection and recognition.}
	\label{fig:overall}
\end{figure*}

Both detection and recognition have evolved through two broad generations of methods. Early approaches relied on handcrafted acoustic features, including tonal features extracted through low-frequency analysis and recording (LOFAR), modulation spectra obtained through detection of envelope modulation on noise (DEMON), and Mel-frequency cepstral coefficients (MFCCs), combined with shallow classifiers for recognition~\cite{barros2022cyclostationarity, hummel2024survey}, while detection typically employed energy detectors or, when suitable signal templates were available, matched filters~\cite{abraham2019underwater}. Although these methods offered a degree of physical interpretability, their performance was often limited in complex acoustic environments. The advent of deep neural networks marked a significant advance, improving recognition accuracy and detection capability by learning discriminative representations directly from time–frequency data~\cite{liu2021crnn,hong2021residual,dong2025passivedetection}. Despite these advances, the underlying decision paradigm has remained largely opaque. Modern systems typically accompany a detection decision or recognition label with a confidence score. Such scores indicate the system’s degree of certainty but offer little insight into the acoustic evidence that supports its conclusion. Moreover, confidence estimates may be poorly calibrated~\cite{guo2017calibration} and do not reveal how individual acoustic cues contribute to a decision. Addressing this limitation requires more than improvements in predictive accuracy. It calls for a fundamental shift in how sensing decisions are formulated and presented. The goal is to move beyond hard labels and scalar confidence scores toward evidence-grounded decisions that explain not only what the system concludes, but also which physical signatures support that conclusion. This capability is critical to establishing calibrated human trust and enabling machine-to-machine verification in distributed sensing systems~\cite{yang2026distributedacoustic,yao2026cfddc}.

Building on this vision, we propose an Underwater Acoustic Physics-aware Description (UPD) framework, a unified detection-and-recognition framework that couples a physics-aware acoustic encoder, as illustrated in Fig. \ref{fig:overall}. The framework is organized around an ordered feature–evidence–decision chain: raw observations are first transformed into physically meaningful tokens anchored to identifiable acoustic structures(tonal lines, envelope modulations,
machinery components), the shared LLM reasoning
core then identifies and states the acoustic evidence in light of textual physical priors, and final judgments are produced conditioned on the stated evidence rather than directly on the features, distinguishing the framework from post-hoc explanation schemes. Its input layer is deliberately open: representative inputs include, but are not limited to, waveform-level observations from single or multiple hydrophones possibly behind physical front ends such as beamforming, environmental and geographic information as imagery or structured parameters, and textual physical priors, and new sources are incorporated simply by adding encoders or extending the textual context, without architectural modification. Detection and recognition are two instantiations of the same chain, sharing the semantic core while differing in operating conditions, input nature, depth of the physical front end, and the role of the environment. In the present Part~I, we instantiate this general framework for recognition through an Underwater Acoustic Physics-aware Transformer Encoder(UPTE), which learns task-disentangled physical representations through four physics-oriented acoustic tasks; the LLM then generates a structured description that first presents acoustic evidence, including operating state, shaft-frequency range and its temporal trend, blade-passing frequency, machinery-related spectral lines, and observation conditions, and subsequently identifies the vessel based on this evidence. This evidence-first formulation also allows the model to characterize unknown-class samples through their acoustic attributes without forcing a decision among the predefined vessel categories. To overcome data scarcity, we further develop a physics-parameterized data engine that synthesizes vessel-radiated noise with fully known source and channel parameters, enabling the construction of two complementary datasets for dual-granularity supervision: a physical-attribute dataset with discrete labels and a structured-description dataset with evidence-grounded textual annotations.

Given this generality, the two instantiations are developed in two companion papers, following the detection–recognition chain in the direction of increasing signal-condition difficulty. 

\emph{Part I}, the present paper, establishes the framework and validates it through the recognition instantiation described above, which operates on detected, preprocessed contacts where signal conditions and physical priors are comparatively well defined, thereby offering the most direct demonstration of the proposed paradigm. 

\emph{Part II} advances upstream to the detection instantiation under low-SNR conditions, which operates closer to the raw acoustic field and demands a deeper integration of propagation physics into the framework~\cite{li2026part2}.

The main contributions of the present paper (Part I) are summarized as follows:
\begin{itemize}
\item We propose the UPD framework, a unified physics-aware approach to underwater acoustic target detection and recognition that, to our knowledge, is the first to couple a physics-aware acoustic encoder with an LLM to produce structured, evidence-bearing descriptions of vessel-radiated noise; its Part~I instantiation targets recognition, advancing it from hard labels toward interpretable decisions.

\item We design the UPTE, which tokenizes complementary acoustic representations and learns task-disentangled physical representations through four physics-oriented tasks, endowing acoustic tokens with explicit physical meaning.

\item We develop a physics-parameterized data engine that constructs two complementary datasets for dual-granularity supervision, pairing each synthesized signal with discrete physical-attribute labels in one dataset and structured, evidence-grounded textual descriptions in the other, enabling physically grounded training and field-wise evaluation.

\item Extensive experiments on simulated data demonstrate accurate recognition of four known vessel types, along with physically grounded descriptions of unknown-class samples, validating the feasibility of evidence-grounded underwater acoustic recognition.
\end{itemize}

The remainder of this paper is organized as follows. Section 2 reviews related work. Section 3 presents the UPD framework. Section 4 details its recognition instantiation. Section 5 presents the physics-parameterized data engine. Section 6 reports experimental results, followed by discussion and conclusions in Section 7.

\section{Related Work}

\subsection{Underwater Acoustic Target Recognition}

Underwater acoustic target recognition has long relied on handcrafted representations of ship-radiated noise. Classical acoustic analysis methods extract characteristic spectral and modulation patterns from received signals. Representative techniques include LOFAR, which reveals narrowband tonal structures~\cite{chen2021lofar}, and DEMON, which extracts modulation components associated with propeller rotation~\cite{guo2024modulation}. The resulting shaft-frequency lines, blade-rate harmonics, and machinery-related components provide physically meaningful cues for identifying targets and interpreting their operating states. Other feature-based approaches use Mel-frequency cepstral coefficients (MFCCs) to summarize spectral characteristics and combine them with statistical classifiers such as Gaussian mixture models~\cite{korany2012classification}. These methods separate expert-designed feature extraction from the subsequent recognition stage. However, extracting stable features remains challenging when weak tonal components are obscured by noise or when signal characteristics vary over time. Their dependence on expert-defined features also limits scalability and adaptability to unfamiliar targets and operating conditions. 

The past decade has witnessed a shift toward data-driven approaches. Convolutional and recurrent neural networks operating on time–frequency representations have substantially improved classification accuracy on public vessel-noise datasets such as ShipsEar and DeepShip~\cite{liu2021crnn,hong2021residual,irfan2021deepship}. Subsequent studies introduce attention mechanisms and residual architectures to emphasize informative features, as exemplified by the attention-based residual network of Li et al.~\cite{li2022aresnet}. Transformer-based models extend the modeling of dependencies across time--frequency tokens. MHT-UATR-Transformer~\cite{feng2024masking} further combines hierarchical tokenization with masked reconstruction for self-supervised representation learning. Researchers have also explored data augmentation~\cite{liu2021crnn}, transfer learning~\cite{li2022aresnet}, and few-shot adaptation~\cite{fu2025mfuatr} to train recognition models with limited labeled data. Despite these advances in representation and learning, the output  typically remains a categorical label, occasionally accompanied by a scalar confidence score. Such outputs state neither which acoustic structures support the decision nor how environmental factors have shaped the observation. Improved classification performance therefore does not, by itself, provide an explicit account of the physical evidence underlying recognition.

More recent studies incorporate physical characteristics or linguistic knowledge into the recognition process. Xie et al. propose DEMONet~\cite{xie2026demonet}, which uses DEMON-derived characteristics to route signals among expert networks. In language-guided recognition, Xie et al. introduce UART~\cite{xie2022uart}, a tri-modal contrastive framework that aligns audio, spectrograms, and templated descriptions. Fu et al. develop MF-UATR~\cite{fu2025mfuatr}, which combines pretrained language-audio representations with multi-feature contrastive learning. Cao et al. propose ASE-CLAP~\cite{cao2026aseclap}, which uses an LLM to construct acoustic-semantic descriptions and enrich class prototypes for similarity-based classification. These methods demonstrate the value of physical and semantic information, but the decision paradigm has remained essentially unchanged: such information primarily improves category prediction, while the model does not generate a structured account of the acoustic evidence for each recording. This limits the basis for human assessment or machine-to-machine verification in operational and distributed sensing scenarios. Advancing underwater acoustic recognition therefore requires a fundamental shift in the decision paradigm, moving beyond categorical labels and confidence scores toward structured descriptions grounded in acoustic evidence. Achieving this shift requires modeling capabilities beyond those of conventional classifier architectures.

\subsection{Large Audio-Language Models and Domain-Specific Adaptation}
Large language models pretrained on massive text corpora have demonstrated strong sequence modeling, knowledge retention, and instruction-following capabilities. These capabilities have been extended to the audio modality through a series of audio-language and large audio-language models (LALMs)~\cite{zhang2023speechgpt,chu2023qwenaudio,tang2024salmonn,kong2024audioflamingo}. A common architecture pairs a pretrained audio encoder~\cite{huang2022audiomae,chen2023beats} with an LLM through a lightweight connector module, such as a learnable projector or a Q-Former, enabling audio-conditioned question answering and description generation. The encoder represents the acoustic observation, while the connector maps its outputs into embeddings that the language model can process alongside textual instructions. SALMONN~\cite{tang2024salmonn}, for example, connects complementary speech and general-audio encoders to an LLM through a window-level Q-Former. Qwen-Audio~\cite{chu2023qwenaudio} supports a broad range of speech, music, and sound-understanding tasks, while Audio Flamingo~\cite{kong2024audioflamingo} extends audio interaction to in-context learning and multi-turn dialogue. GAMA~\cite{ghosh2024gama} further explores reasoning-oriented instruction tuning for complex audio understanding. These developments enable a single model to perform multiple audio-understanding tasks through natural-language instructions and generate descriptions of sound events and their acoustic characteristics.

Parameter-efficient fine-tuning reduces the computational cost of adapting pretrained models. LoRA~\cite{hu2022lora}, for example, keeps pretrained weights frozen and learns low-rank updates, reducing the number of trainable parameters and memory requirements. The availability of labeled data is another consideration in adapting pretrained models to specialized domains. Domain-specific systems such as LLaVA-Med~\cite{li2023llavamed} and GeoChat~\cite{kuckreja2024geochat} adapt general vision-language models to biomedical imaging and remote sensing through domain alignment and instruction tuning. Their results suggest that pretrained knowledge and reasoning capabilities can be reused with targeted supervision, an attractive direction for underwater acoustics, where annotated recordings are scarce. However, generating a convincing explanation does not ensure that it reflects the evidence used to make a prediction~\cite{turpin2023unfaithful,matton2025walk}. For underwater recognition, this motivates grounding generated descriptions in verifiable acoustic evidence.

In addition, existing general-purpose LALMs are primarily designed to understand speech, music, and environmental sounds, and their ability to recognize physically structured underwater signals remains uncertain. A concurrent benchmark, UA-Bench~\cite{sun2026uabench}, evaluated nine open LALMs without task-specific adaptation on synthesized
underwater acoustic signals. Under its gated coarse-to-fine evaluation protocol, the highest leaf-level accuracy was 13.1\%, compared with a random reference of 7.7\%. Four models showed reliable improvements in leaf-level accuracy over a paired silent-audio control, with a maximum improvement of 4.7 percentage points. These findings indicate limitations in the evaluated models' transfer to fine-grained underwater recognition. They motivate exploring physics-aware adaptation to capture the underwater-specific physical characteristics on which vessel recognition depends. The present work develops such an adaptation framework: we couple a physics-aware acoustic encoder with a pretrained LLM to produce fine-grained, evidence-bearing physical descriptions of vessel-radiated noise.

\section{UPD Framework: From-Signals-to-Evidence}
\label{sec:framework}


\begin{figure*}[!t]
	\centering
	\includegraphics[width=0.95\textwidth]{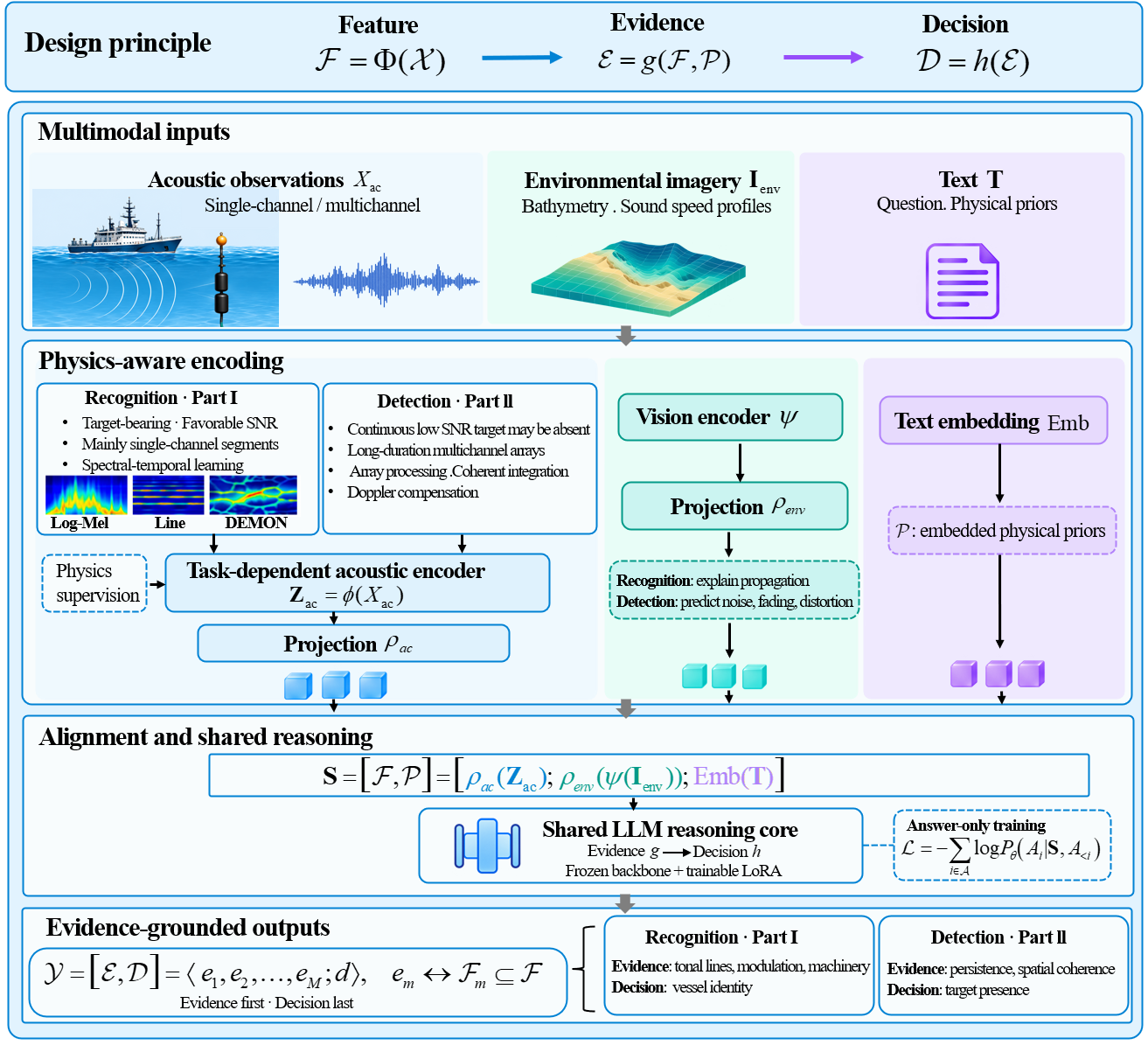}
	\caption{Architecture of the UPD framework based on the feature–evidence–decision chain.}
	\label{fig:schema}
\end{figure*}

Conventional underwater acoustic systems compress an observation into a decision, whether a detection flag or a class label, discarding the intermediate physical reasoning that a human expert would articulate. The proposed framework in contrast makes the intermediate reasoning explicit and organizes the entire processing pipeline around an ordered \emph{feature--evidence--decision} chain, as illustrated in Fig. \ref{fig:schema}
\begin{equation}
\mathcal{F} = \Phi(\mathcal{X}), \qquad
\mathcal{E} = g(\mathcal{F}, \mathcal{P}), \qquad
\mathcal{D} = h(\mathcal{E}),
\label{eq:chain}
\end{equation}
where $\mathcal X$ denotes the available non-textual multimodal observations, $\Phi$ represents the corresponding feature construction process, and $\mathcal F$ denotes the resulting multimodal physical representations. $\mathcal{P}$ represents the embedded representations of textual
physical priors, while $g$ and $h$ abstract the evidence-forming and decision-making processes of the shared LLM reasoning core. The observations are first transformed into physically meaningful representations (\emph{feature}); the reasoning core then identifies and articulates task-relevant physical evidence in light of textual priors (\emph{evidence}); and the final judgment is formulated based on the stated evidence (\emph{decision}). Notably, the decision $\mathcal{D}$ is conditioned on the evidence $\mathcal{E}$ rather than directly on the features $\mathcal{F}$, which distinguishes this chain from post-hoc explanation schemes in which justifications are generated independently of the decision process.

Two further principles guide the design. First, physical grounding is maintained through modality-aware feature construction. Specifically, acoustic representations are anchored to identifiable physical structures, such as tonal lines, envelope modulations, and machinery-related components, rather than opaque activations, allowing the resulting evidence to be verified against the input. Other modalities, such as environmental and geographic information, can provide complementary physical context for interpreting acoustic observations, particularly with respect to propagation effects and observation conditions. Second, heterogeneous information is integrated within a shared LLM input space. Multimodal physical representations and textual physical priors are mapped into the embedding space of an LLM, which serves as the unified reasoning core. Although the framework is instantiated differently for detection and recognition, both tasks follow the same feature–evidence–decision principle and share the same reasoning core.


\subsection{Multimodal Input Layer}
\label{sec:input}

The framework is designed to accommodate the heterogeneous information available in underwater sensing, without prescribing a closed set of input types. Representative inputs include, but are not limited to: (i) \emph{acoustic modality}, comprising waveform-level observations, i.e., the received time series from single or multiple hydrophones, possibly preprocessed by a physical front end such as beamforming; (ii) \emph{vision modality}, encompassing environmental and geographic information, such as sound speed profiles, bathymetric charts, sea state, and seabed properties, expressed either as imagery or as structured parameters; and (iii) \emph{textual modality}, consisting of textual physical priors and expert knowledge expressed in natural language. Whereas conventional architectures often rely on ad-hoc feature engineering to integrate heterogeneous information, the language-modeling interface allows such information to be incorporated into a common input sequence through modality-specific encoders or textual representations, without architectural modification. The input layer is thus extensible: new information sources can be incorporated by adding the corresponding encoders or simply by extending the textual context.

\subsection{Physics-Aware Acoustic Encoding}
\label{sec:encoding}

As the acoustic component of the multimodal feature construction process $\Phi$ in Eq.~\eqref{eq:chain}, the physics-aware encoder $\phi$ maps acoustic observations $X_{\mathrm{ac}}$ into tokens associated with identifiable acoustic structures, providing concrete referents for subsequent evidence statements. Its design follows three principles. First, acoustic observations are transformed into physically meaningful representations through task-appropriate encoding pathways, preserving the acoustic structures relevant to the sensing objective. Second, the encoder is trained on physics-oriented analysis tasks, such as shaft frequency and modulation analysis for recognition and coherent integration and bearing analysis for detection, rather than optimized solely for the final decision. This encourages the learned tokens to capture how a target sounds, rather than merely which label it belongs to. Third, conventional physical processing, including filtering, normalization, beamforming, and Doppler compensation, is retained as a parameterized front end wherever appropriate. The degree of physical preprocessing varies between detection and recognition, reflecting their different signal conditions and processing requirements. (Section~\ref{sec:unified}).

\subsection{LLM Reasoning Core and Cross-Modal Integration}
\label{sec:core}

All input modalities are processed by their respective encoders and organized into a unified token sequence for the LLM reasoning core

\begin{equation}
\mathbf{S} = \big[\, \rho_{ac}(\mathbf{Z}_{\mathrm{ac}});\ 
\rho_{env}(\psi(\mathbf{I}_{\mathrm{env}}));\ 
\mathrm{Emb}(\mathbf{T}) \,\big],
\label{eq:sequence}
\end{equation}
where $\mathbf{Z}_{\mathrm{ac}}=\phi(X_{\mathrm{ac}})$ denotes acoustic tokens produced by the acoustic encoder, $\rho(\cdot)_{ac}$ and $\rho(\cdot)_{env}$ are modality-specific projection modules, $\mathbf{I}_{\mathrm{env}}$ environmental imagery with its encoder $\psi$, and $\mathrm{Emb}(\mathbf{T})$ the embeddings of textual tokens including queries, instructions, and physical priors. The projection modules provide modality-specific interfaces to the shared embedding space. The concatenated sequence $\mathbf{S}$ is processed by an LLM, which acts as the shared reasoning core: it aggregates evidence across representations and priors, and generates responses in an ordered evidence-then-decision manner. The pretrained parameters are kept frozen and adapted through low-rank modules, so that the pretrained knowledge and instruction-following capability are preserved while the learning burden is confined to the acoustic-to-semantic mapping. Training supervises only the answer portion of each sequence
\begin{equation}
\mathcal{L} = - \sum_{i \in \mathcal{A}} 
\log P_{\theta}\!\left( A_i \,\middle|\, \mathbf{S}, A_{<i} \right),
\label{eq:loss}
\end{equation}
where $\mathcal{A}$ denotes the answer token positions, and losses over queries, instructions, and input portions are masked out. This design directs the learning signal toward evidence-based reasoning: extracting relevant evidence from acoustic features and using it to support subsequent decisions, rather than reproducing the input. By sharing a common reasoning core across tasks, the framework also provides a flexible query interface through which a single model addresses vessel identification, physical attribute estimation, and target detection, without relying on separate task-specific output heads.

\subsection{Evidence-Grounded Output}
\label{sec:output}

The output of the framework is a structured description rather than a bare decision. Formally, each response is organized as

\begin{equation}
\mathcal{Y} = \langle \ e_1, e_2, \dots, e_M; d \rangle, 
\qquad e_m \leftrightarrow \mathcal{F}_m \subseteq \mathcal{F},
\label{eq:output}
\end{equation}
where $\mathcal{Y}$ denotes the complete structured output, each evidence item $e_m$ is linked to a verifiable subset $\mathcal{F}_m$ of the multimodal physical representations and $d$ represents the final decision. Evidence statements may refer, for example, to tonal-line locations, modulation patterns, and estimated observation conditions, as well as to missing components that are attributable to the propagation environment rather than to the absence of the target. A fixed field order is imposed on the description, which enforces the evidence-then-decision ordering during generation, renders every attribute individually parseable for evaluation, and allows each reported statement to be independently verified against the input. In this sense, the framework replaces a single opaque score with an inspectable chain whose correctness can be checked link by link.

\subsection{A Unified View of Detection and Recognition}
\label{sec:unified}

Detection and recognition have traditionally been addressed as separate problems using distinct architectures. Within the proposed framework, however, they are two instantiations of a common feature--evidence--decision chain, sharing the same semantic core but differing in four key respects. First, their operating conditions differ. Recognition typically operates on target-bearing segments, often identified through prior detection, where the signal-to-noise ratio is relatively favorable and the relevant physical priors are well established. Detection, by contrast, must operate on continuous observations in which the target, if present, may be deeply buried in noise. 

Second, their input requirements differ. Recognition primarily processes single-channel, target-centered recordings, whereas detection must accommodate multichannel array signals over extended observation periods and provide spatial information, such as target bearing, alongside target-presence decisions.

Third, the extent of physical front-end integration differs. Detection relies more heavily on parameterized physical processing, including array processing, coherent integration, and Doppler compensation, applied before or within the encoder. Recognition, in contrast, places greater emphasis on learning representations of spectral and temporal structure. 

Fourth, environmental information serves different purposes. In recognition, bathymetry and sound speed profiles are used to model propagation effects during data generation, introducing realistic channel variations into the acoustic observations. In detection, they also support predictive modeling of the noise field, fading, and channel-induced distortion. Anticipating these effects is important for identifying potential target signatures against background interference and improving bearing estimation. 

This paper (Part I) establishes the framework and validates its recognition instantiation, where target-bearing recordings allow the relationship between acoustic features and physical attributes to be examined directly. Part II extends the framework to low-SNR detection, with particular emphasis on multichannel processing and environmental-based prediction.

\begin{figure*}[t]
	\centering
	\includegraphics[width=1\textwidth]{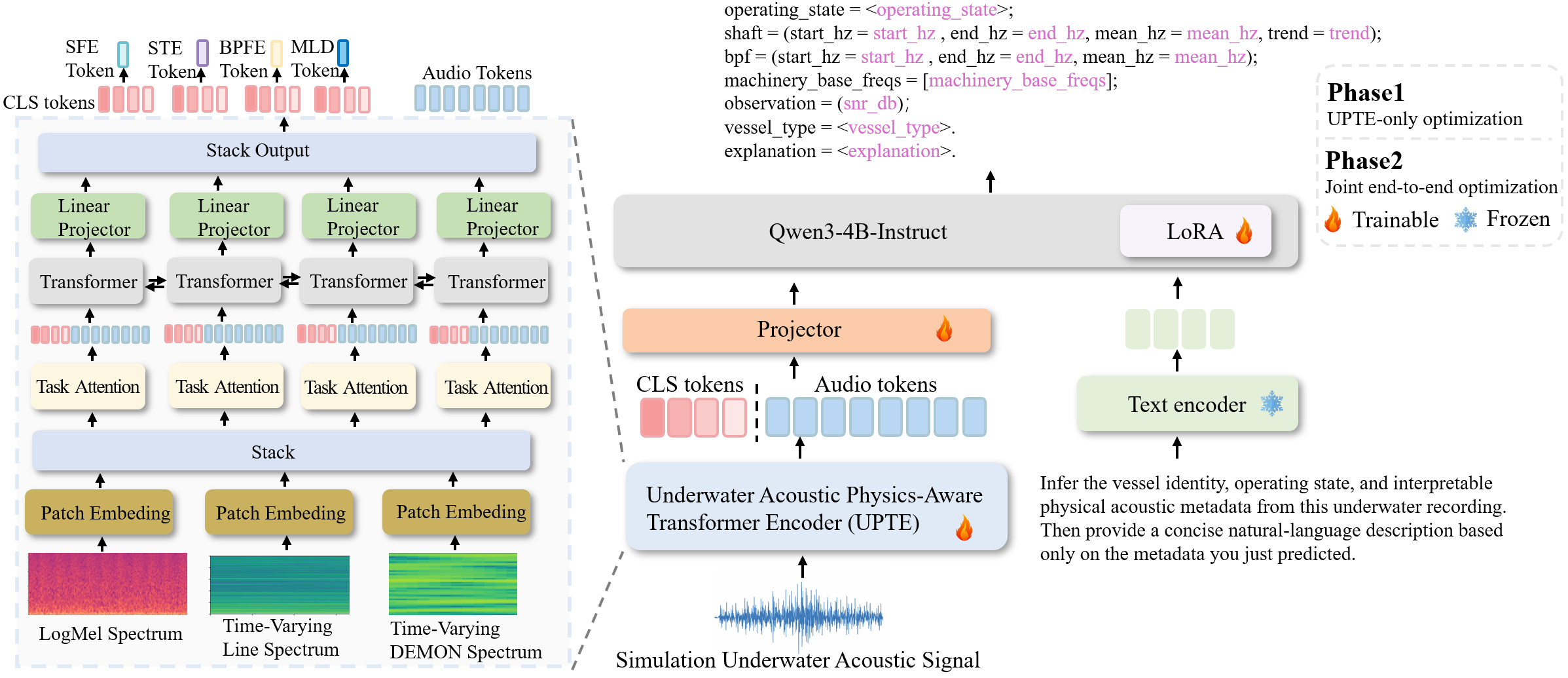}
	\caption{Architecture of the recognition instantiation of the UPD framework, integrating the physics-aware UPTE with a pretrained LLM. The training process consists of two phases: only the UPTE is trained in Phase 1, while the entire framework is jointly trained in Phase 2
		}
	\label{fig:structured_output}
\end{figure*}

\section{Recognition Instantiation: Underwater Acoustic Physics-aware Transformer}
\label{sec:instantiation}

In the remainder of this paper, we instantiate the UPD framework for underwater acoustic target recognition. Following the framework design in Section~\ref{sec:framework}, the recognition instantiation independently tokenizes complementary representations (Section~4.1), the encoder is supervised by physics-oriented analysis tasks rather than vessel type alone (Sections~4.2--4.3), and the resulting tokens are fed into the reasoning core under an answer-only objective (Section~4.4). The resulting recognition-specific encoder is called Underwater Acoustic Physics-aware Transformer Encoder(UPTE), and the complete recognition system couples UPTE with a pretrained LLM, as illustrated in Fig.\ref{fig:structured_output}.


\subsection{Acoustic Representation and Tokenization}
\label{sec:tokenization}

\subsubsection{Acoustic Representations}

To capture complementary acoustic characteristics, three time–frequency representations with distinct physical meanings and resolutions are extracted from each received signal $x(t)$. The Log-Mel spectrogram captures the overall energy distribution
\begin{equation}
X_{\mathrm{mel}} = \log (\left| \mathrm{STFT}\{x(t)\} \right|^2 W_{\mathrm{mel}}),
\label{eq:mel}
\end{equation}
where $W_{\mathrm{mel}}$ denotes the Mel filterbank. To resolve narrowband rotating-machinery components, a time-varying line spectrum(Line-TF) is computed from a resampled waveform using a long-window STFT
\begin{equation}
X_{\mathrm{line}}(\tau, f) = \log (\left| \mathrm{STFT}\{x(t)\}(\tau, f) \right|^2),
\label{eq:line}
\end{equation}
which preserves the trajectories of tonal components over time. The time-varying DEMON spectrogram characterizes the temporal evolution of low-frequency envelope modulation: after band-pass filtering, the analytic envelope
\begin{equation}
e(t) = \left| x_b(t) + j\,\mathcal{H}\{x_b(t)\} \right|
\label{eq:envelope}
\end{equation}
is computed, where $\mathcal{H}\{\cdot\}$ denotes the Hilbert transform, and its modulation spectrum is obtained via a second STFT
\begin{equation}
X_{\mathrm{DEMON}}(\tau, f_m) = \log \left| \mathrm{STFT}\{e(t)\}(\tau, f_m) \right|.
\label{eq:demon}
\end{equation}
The three representations thus expose complementary physical evidence: broadband energy patterns, tonal-line structures, and propeller-related modulations.

\subsubsection{Independent tokenization}
Because the three acoustic representations differ in their physical meanings and time-frequency resolutions, directly concatenating them in the raw feature domain is inappropriate. We therefore tokenize each representation independently
\begin{equation}
	\mathbf{Z}_k = E_k(\mathbf{X}_k),
	\qquad
	k\in\{\mathrm{mel},\mathrm{line},\mathrm{DEMON}\}
\end{equation}
where $E_k$ denotes the representation-specific patch embedding operator. Different patching strategies are adopted to accommodate their distinct time-frequency structures. Each token stream is further augmented with an independent positional embedding and a learned feature-type embedding, enabling the transformer to preserve the identity of different physical representations after token-level fusion. 

\subsection{Task-Specific Feature Attention}
\label{sec:attention}

To learn physically grounded acoustic representations, UPTE is organized around four physics-oriented analysis tasks: Shaft Frequency Estimation (SFE), Shaft Trend Estimation (STE), Blade-Passing Frequency Estimation (BPFE), and Machinery Line Detection (MLD), whose physical labels are derived from the simulation parameters (Section~5). The three representations contribute unequally to these tasks. A task-specific feature attention module is therefore employed to adaptively weight the representations for each task.

Each token stream $\mathbf{Z}_k \in \mathbb{R}^{B \times P_k \times D}$ is first summarized by combining average and max pooling over the token dimension
\begin{equation}
\mathbf{u}_k = \big[\, \mathrm{AvgPool}(\mathbf{Z}_k);\ \mathrm{MaxPool}(\mathbf{Z}_k) \,\big] \in \mathbb{R}^{B \times 1 \times 2D},
\label{eq:pool}
\end{equation}
where $B$, $P_k$, and $D$ denote the batch size, the number of tokens in the $k^\text{th}$ stream acoustic representations, and the token embedding dimension, respectively. Among them, average pooling captures the overall acoustic characteristics of all streams, whereas max pooling preserves salient local responses, such as prominent tonal components and machinery-related spectral structures. For each task, an independent attention network estimates the relevance of the three representations from the stacked summary $\mathbf{U} \in \mathbb{R}^{B \times 3 \times 2D}$
\begin{equation}
\mathcal{M} = \mathbf{W}_2\,\mathrm{GELU}\!\left( \mathbf{W}_1\,\mathrm{LN}(\mathbf{U}) + \mathbf{b}_1 \right) + \mathbf{b}_2,
\label{eq:attn}
\end{equation}
followed by Softmax normalization across the representation dimension, where $\operatorname{GELU}(\cdot)$ denotes the GeLU activation function, $\operatorname{LN}(\cdot)$ denotes layer normalization, and $\mathbf{W}_{1}$, $\mathbf{W}_{2}$, $\mathbf{b}_{1}$, and $\mathbf{b}_{2}$ denote the weights and bias terms of the two fully connected layers. A minimum contribution is retained for each stream, preventing potentially complementary physical information from being completely suppressed. Beyond adaptivity, the learned attention coefficients provide an interpretable measure of how much each physical representation contributes to each analysis task, which we examine in Section~6.

\subsection{Task-Disentangled Branches}
\label{sec:branches}

To prevent the physical semantics of different tasks from interfering with one another, the information flow is separated into four task-specific branches. In each branch, the shared acoustic tokens are weighted by the corresponding task-specific attention module; four learnable CLS tokens are then prepended to the weighted sequence and processed by a shared Transformer encoder consisting of 12 blocks, initialized with pretrained AudioMAE weights\cite{huang2022audiomae}. The resulting token representations are projected into a task-specific subspace through an independent linear projection layer, which reduces the embedding dimension from $D$ to $D/4$; since all branches preserve the same token ordering, their projected representations are concatenated along the feature dimension, yielding unified token sequences with the original dimension $D$. This design allows the four physical semantics to be disentangled in dedicated subspaces while keeping the backbone shared.

During UPTE training, the projected CLS representations of each branch are averaged and fed into the corresponding prediction head. SFE, STE, and BPFE are formulated as single-label classification tasks and optimized using cross-entropy loss
\begin{equation}
\mathcal{L}q = -\frac{1}{B}\sum_{i=1}^{B}\log p_{i,y_i}^{(q)},
\qquad q \in {\mathrm{SFE},\mathrm{STE},\mathrm{BPFE}},
\label{eq:taskloss}
\end{equation}
where $p_{i,y_i}^{(q)}$ denotes the predicted probability of the ground-truth class $y_i$ for sample $i$ in task $q$. Since multiple machinery components may coexist within one recording, MLD is formulated as multi-label classification. We employ a combined soft-label and peak-coverage loss
\begin{equation}
	\begin{aligned}
		\mathcal{L}_{\mathrm{MLD}}
		&= \lambda_{\mathrm{soft}}\mathcal{L}_{\mathrm{soft}}
		+ \lambda_{\mathrm{peak}}\mathcal{L}_{\mathrm{peak}},\\
		\mathcal{L}_{\mathrm{soft}}
		&= -\sum_{k=1}^{K}\tilde{y}_k\log p_k,\\
		\mathcal{L}_{\mathrm{peak}}
		&= -\frac{1}{N}\sum_{g\in\mathcal G}
		\log\sum_{|k-g|\leq1}p_k.
	\end{aligned}
	\label{eq:uptloss}
\end{equation}
where $\lambda_{\mathrm{soft}}$ and $\lambda_{\mathrm{peak}}$ denote the MLD loss weights, $p_k$ is the predicted probability of frequency bin $k$, $\tilde y_k$ is the normalized target obtained through neighboring-bin smoothing, and $\mathcal G$ denotes the set of $N$ ground-truth machinery-frequency bins, used to encourage the detection of all machinery peaks. The overall objective of UPTE is a weighted combination
\begin{equation}
	\begin{aligned}
		\mathcal{L}_{\mathrm{UPTE}}
		=&\ \lambda_{\mathrm{SFE}}\mathcal{L}_{\mathrm{SFE}}
		+\lambda_{\mathrm{STE}}\mathcal{L}_{\mathrm{STE}}\\
		&+\lambda_{\mathrm{BPFE}}\mathcal{L}_{\mathrm{BPFE}}
		+\lambda_{\mathrm{MLD}}\mathcal{L}_{\mathrm{MLD}}.
	\end{aligned}
	\label{eq:uptloss}
\end{equation}
where $\lambda_{\mathrm{SFE}}$, $\lambda_{\mathrm{STE}}$, $\lambda_{\mathrm{BPFE}}$, and $\lambda_{\mathrm{MLD}}$ denote the task weights. The task-specific attention modules, projection layers, and prediction heads are optimized by their corresponding task losses, while the shared patch embedding modules and Transformer encoder are jointly optimized through the combined objective. This design enables UPTE to learn task-specialized representations while exploiting shared acoustic information across the four physical analysis tasks. By grounding representation learning in explicit physical tasks rather than vessel type, UPTE captures how a vessel sounds—its propulsion and machinery behavior.

\subsection{Answer-Only Adaptation}
\label{sec:alignment}

The unified token representations produced by UPTE, comprising a set of shared acoustic tokens and four task-specific CLS tokens, are projected into the embedding space of a pretrained LLM (Qwen3-4B-Instruct) through a learnable projection module, and are concatenated with the embeddings of the textual query as in Eq.~\ref{eq:sequence}. Following the adaptation strategy of Section~\ref{sec:core}, LoRA modules are inserted into the LLM, while all remaining pretrained parameters are kept frozen; only the newly introduced UPTE modules, the acoustic projection, and the LoRA adapters are optimized. This configuration preserves the pretrained knowledge and reasoning capability of the LLM and confines the learning burden to the acoustic-to-semantic mapping, which is essential given the limited scale of underwater acoustic data.

Following Eq.~\ref{eq:loss}, training loss is applied only to the answer tokens of the response. The two-stage training protocol first optimizes UPTE using Eq.~\ref{eq:uptloss} and then optimizes the complete framework using Eq.~\ref{eq:loss}, as detailed in Section~6.1.

\section{Physics-Parameterized Data Engine}
\label{sec:data}

\begin{figure*}[t]
	\centering
	\includegraphics[width=1.0\textwidth]{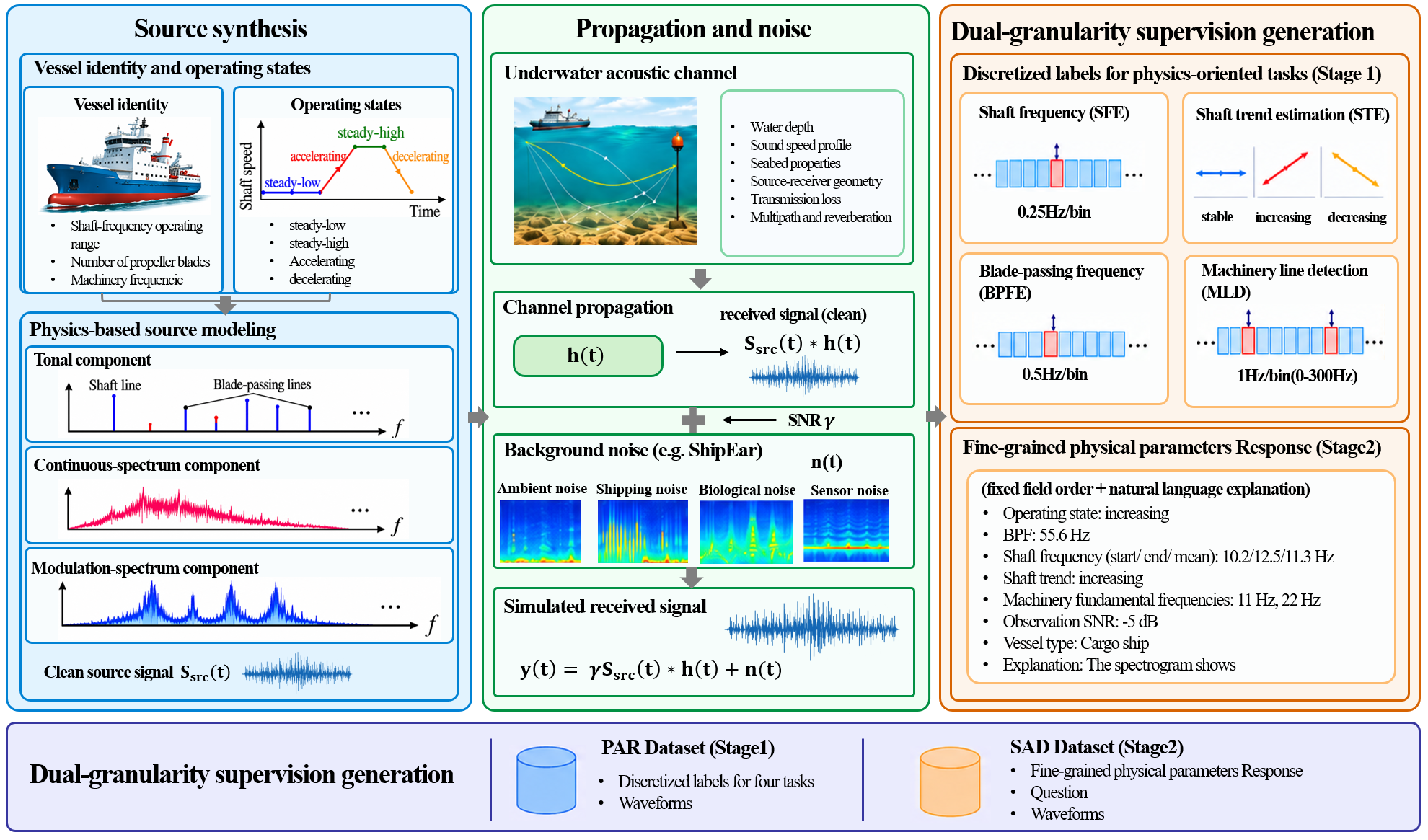}
	\caption{Architecture of the physics-parameterized data engine for dual-granularity supervision.}
	\label{fig:data_generate}
\end{figure*}

Underwater acoustic recordings with physical annotations remain scarce, as the source parameters and propagation conditions of real-world recordings are often unknown. This makes it difficult to obtain reliable physical-attribute labels for evidence-grounded training. We therefore construct a \emph{physics-parameterized data engine} that generates underwater acoustic recordings through a parameterized synthesis pipeline. Each recording is generated from explicitly sampled parameters governing vessel identity, operating state, source signal composition, propagation conditions, and background noise, ensuring the recording is \emph{traceable} to the conditions under which it was generated. This traceability provides ground truth for the physics-oriented tasks of UPTE and supports evidence-grounded supervision of the structured descriptions generated by UPD. As illustrated in Fig.~\ref{fig:data_generate}, the engine is organized into source synthesis (Section~5.1), propagation and noise (Section~5.2), and dual-granularity supervision generation (Section~5.3).

\subsection{Source Signal Synthesis}
\label{sec:source}

\subsubsection{Vessel identity and operating states}

Each simulated vessel is assigned a fixed set of identity parameters, comprising the shaft-frequency $f_{s}$ range, number of propeller blades $Z$, and two machinery fundamental frequencies $f_{mach}=(f_{m,1},f_{m,2})$. For each recording, the instantaneous shaft frequency is defined as
$$
f_s(t)=f_{s,0}+\Delta f_s t,\qquad 0\leq t\leq T,
$$
where $f_{s,0}$ is the starting shaft-frequency, $\Delta f_s$ is the frequency change rate, and $T$ is the recording duration in seconds. Four operating states are considered: steady low speed and steady high speed, with $\Delta f_s=0$ and $f_{s,0}$ sampled from the lower and upper portions of the operating range, respectively; and accelerating and decelerating, characterized by positive and negative $\Delta f_s$, respectively.

\begin{figure*}[htbp]
	\centering
	\includegraphics[width=1.0\textwidth]{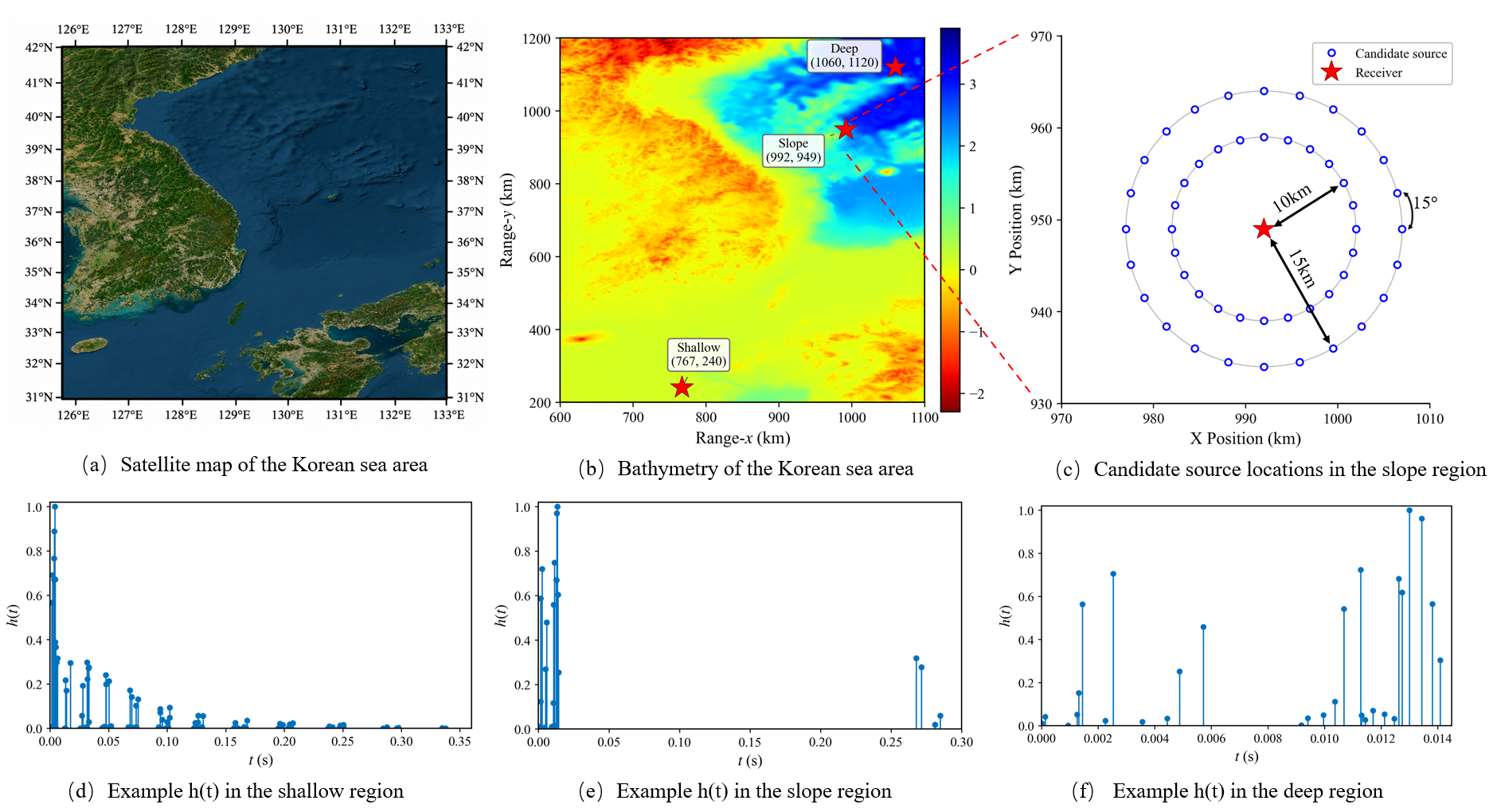}
	\caption{Selected Korean sea regions and their representative channel impulse responses. The examples in (d)–(f) are obtained at a range of 15 km and an azimuth of $195^\circ$.}
	\label{fig:korean_bathymetry}
\end{figure*}
\subsubsection{Source signal composition}
Following the classical understanding of ship-radiated noise, the source signal is modeled as a weighted sum of three acoustically distinct components: tonal, continuous-spectrum, and modulation-spectrum
\begin{equation}
	s_{\mathrm{src}}(t) = w_{\mathrm{tonal}}\,s_{\mathrm{tonal}}(t) 
	+ w_{\mathrm{cont}}\,s_{\mathrm{cont}}(t)
	+ w_{\mathrm{mod}}\,s_{\mathrm{mod}}(t),
	\label{eq:source}
\end{equation}
where $w_{\mathrm{tonal}}$, $w_{\mathrm{mod}}$, and $w_{\mathrm{cont}}$ are the corresponding weights.

\textbf{Tonal component} consists of shaft frequency harmonics associated with propeller shaft rotation, BPF harmonics induced by periodic blade interactions, and machinery related spectral lines generated by onboard mechanical vibrations, which is expressed as
\begin{equation}
	s_{\mathrm{tonal}}(t)
	=
	s_{\mathrm{shaft}}(t)
	+
	s_{\mathrm{BPF}}(t)
	+
	s_{\mathrm{mach}}(t).
\end{equation}

Since all three contributions exhibit harmonic structures, they are modeled using a unified formulation
\begin{equation}
	\begin{aligned}
		s_q(t) &= \sum_{j=1}^{J_q}\sum_{k=1}^{H_{q,j}}
		\frac{A_{q,j}}{k^{\alpha_{q,j}}}
		\cos\left(k\phi_{q,j}(t)+\varphi_{q,j,k}\right),\\
		&\quad q\in\{\mathrm{shaft},\mathrm{BPF},\mathrm{mach}\}.
	\end{aligned}
\end{equation}
where $J_q$ is the number of fundamental frequencies, $H_{q,j}$ is the maximum harmonic order, $A_{q,j}$ is the amplitude scaling factor, $\alpha_{q,j}$ is the harmonic decay exponent, and $\varphi_{q,j,k}$ is the random initial phase. Specifically, $J_{\mathrm{shaft}} = J_{\mathrm{BPF}}=1$ and $J_{\mathrm{mach}} = 2$.

The accumulated phase $\phi_{q,j}(t)$ and the corresponding instantaneous fundamental frequency $f_{q,j}(t)$ are defined as
\begin{equation}
	\begin{aligned}
		\phi_{q,j}(t)
		&=2\pi\int_0^t f_{q,j}(\tau)\,\mathrm{d}\tau,\\[6pt]
		f_{q,j}(t)
		&=
		\begin{cases}
			f_s(t), & q=\mathrm{shaft},\\
			Zf_s(t), & q=\mathrm{BPF},\\
			f_{m,j}, & q=\mathrm{mach}.
		\end{cases}
	\end{aligned}
	\label{eq:phase_frequency}
\end{equation}

\textbf{Continuous-spectrum component} is synthesized using an autoregressive (AR) model driven by Gaussian white noise. Its spectral shape is specified by the target autocorrelation sequence
\begin{equation}
	R[k]
	=
	\cos\left(
	2\pi k\frac{f_c}{f_{\mathrm{sr}}}
	\right)
	\exp\left(
	-2\pi k\frac{f_b}{f_{\mathrm{sr}}}
	\right),
	\qquad
	k = 0,1,...,p.
\end{equation}
where $p$ is the AR order, $f_{\mathrm{sr}}$ is the sampling frequency, while $f_c$ and $f_b$ control the spectral center and bandwidth, respectively. The continuous-spectrum signal is then generated as
\begin{equation}
	s_{\mathrm{cont}}[n]
	=
	\sum_{i=1}^{p}a_i s_{\mathrm{cont}}[n-i]
	+
	\epsilon[n],
\end{equation}
where $\epsilon[n]$ is Gaussian white noise, $a_i$ denotes AR coefficients obtained from $R[k]$ by solving the Yule--Walker equations.

\textbf{Modulation-spectrum component} arises from periodic amplitude fluctuations of broadband noise associated with propeller rotation. As successive propeller blades generate periodic pressure disturbances, a portion of the broadband noise undergoes amplitude modulation. Thus the resulting modulated signal is modeled as
\begin{equation}
	s_{\mathrm{mod}}(t)
	=
	p(t)s_{\mathrm{cont}}(t),
\end{equation}
where $p(t)$ is the modulation envelope. To capture the localized pressure disturbances associated with successive blade passages, the envelope is constructed by superposing Gaussian pulses
\begin{equation}
	p(t)=\sum_{i=1}^{Z} p_i(t),
	\qquad
	p_i(t)
	=
	\frac{\xi}{\sqrt{2\pi}}
	\exp\left[
	-\frac{(t-t_i)^2}{2\sigma_i^2}
	\right],
\end{equation}
where $\xi$ controls the pulse amplitude, and $t_i$ is the occurrence time of the $i^\text{th}$ blade-passing event. To account for variations in propeller rotational speed, the pulse width $\sigma_i$ is adjusted according to the instantaneous BPF
\begin{equation}
	\sigma_i
	=
	\frac{\eta}{Zf_s(t_i)},
\end{equation}
where $\eta$ controls the relative pulse width.

\subsection{Propagation and Background Noise}
\label{sec:channel}

To generate received signals under controlled yet realistic conditions, source signals are processed through multipath propagation models using BELLHOP-based ray tracing and are then combined with segments of \emph{real ocean background noise}. The received signal is
\begin{equation}
	y(t) = \gamma\, h(t) * s_{\mathrm{src}}(t) + n(t),
	\label{eq:received}
\end{equation}
where $h(t)$ denotes the normalized channel impulse response, $n(t)$ is a randomly selected segment of recorded ocean noise, and $\gamma$ is a scaling factor determined by the target signal-to-noise ratio. Different noise segments, SNR levels, and channel realizations can be sampled to generate diverse observations of the same source signal.

For propagation modeling, three representative regions in Korean coastal waters are selected, covering shallow-water, sloping-bottom, and deep-water environments with distinct bathymetric characteristics, as shown in Fig.~\ref{fig:korean_bathymetry}(a)(b). In each region, the receiver position is fixed, while source positions are sampled at ranges of 10 and 15 km across 24 azimuths with $15^\circ$ spacing, as illustrated in Fig.~\ref{fig:korean_bathymetry}(c). For each channel realization, one candidate source location is selected. The source and receiver depths are set to 10 and 30 m, respectively. The corresponding channel impulse responses are constructed from the multipath arrival amplitudes and delays simulated by BELLHOP3D, with bathymetry and sound speed profiles (SSPs) as environmental inputs. Representative channel impulse responses are presented in Fig.~\ref{fig:korean_bathymetry}(d)-(f).

\subsection{Dual-Granularity Supervision}
\label{sec:supervision}

Since the synthesis parameters are known, each recording is paired with two complementary forms of supervision.

\subsubsection{Physics-aware acoustic representation(PAR) Dataset}

The PAR dataset provides discrete labels for the four physics-oriented tasks of UPTE. The starting shaft frequency, BPF, and machinery fundamental frequencies are quantized into uniform bins of 0.25~Hz, 0.5~Hz, and 1~Hz, respectively, with machinery frequencies treated as multi-label targets. The four operating states are mapped to three shaft-frequency trend classes: stable, increasing, and decreasing. These label designs provide explicit supervision while reducing sensitivity to small frequency variations.

\subsubsection{Structured acoustic descriptions(SAD) Dataset}

The SAD dataset consists of $\langle$audio, question, response$\rangle$ triplets, where each recording is paired with a fixed question and a structured response. The response contains physical attributes in a fixed field order, including operating state, shaft-frequency start/end/mean values and trend, BPF values, machinery fundamental frequencies, observation SNR, and vessel type, followed by a natural-language explanation generated by an off-the-shelf pretrained LLM based on these attributes. Unlike the discrete labels in PAR, frequency values and SNR in SAD are specified to 0.01~Hz/dB, preserving finer-grained physical information. The structured physical attributes follow a canonical format for consistency, while the natural-language explanations retain flexibility in expression.

\subsection{Dataset Construction}
\label{sec:datastats}

\begin{table}[t]
	\centering
	\caption{Vessel-specific identity parameters}
	\small
	\begin{tabular}{lccc}
		\toprule
		Vessel & Blades & $f_{s}$ (Hz) 
		& $f_m$ (Hz) \\
		\midrule
		
		Cargo ships
		& 3,4,5 & 1.3-2.5 & 60-80, 145-175 \\
		
		Passenger ships
		& 3,4,5 & 1.6-2.75 & 90-110, 175-205 \\
		
		Fishing vessels
		& 3,4,5 & 2-3.2 & 55-75, 115-145 \\
		
		Motorboats
		& 5,6 & 1.4- 2.4 & 105-125, 220-250 \\
		
		Outliers
		& 3,4,5,6 & 1.25- 3.25 & $\substack{
			f_{m,1},f_{m,2}\in[10,300]\\
			|f_{m,1}-f_{m,2}|\geq 5
		}$ \\
		
		\bottomrule
	\end{tabular}
	\label{tab:vessel_parameters}
\end{table}

\begin{table}[t]
	\centering
	\caption{Shared source-generation parameters and their distributions.}
	\label{tab:source_params}
	
	\small
	\renewcommand{\arraystretch}{1.2}
	\setlength{\tabcolsep}{5pt}
	
	\begin{tabularx}{0.95\columnwidth}
		{@{}l>{\raggedleft\arraybackslash}Xl@{}}
		\toprule
		Param & Setting / Range & Distribution \\
		\midrule
		
		$f_{\mathrm{sr}}$
		& $32000$ & Fixed \\
		
		$T$
		& $8$ & Fixed \\
		
		$w_{\mathrm{tonal}}$
		& $0.50\sim0.60$ & Uniform \\
		
		$w_{\mathrm{cont}}$
		& $0.30\sim0.40$ & Uniform \\
		
		$w_{\mathrm{mod}}$
		& $1-w_{\mathrm{tonal}}-w_{\mathrm{cont}}$
		& Derived \\
		
		$A_{\mathrm{shaft}}$
		& $0.32\sim0.36$ & Uniform \\
		
		$H_{\mathrm{shaft}}$
		& $20$ & Fixed \\
		
		$\alpha_{\mathrm{shaft}}$
		& $1.00\sim1.20$ & Uniform \\
		
		$A_{\mathrm{BPF}}$
		& $0.16\sim0.20$ & Uniform \\
		
		$H_{\mathrm{BPF}}$
		& $8$ & Fixed \\
		
		$\alpha_{\mathrm{BPF}}$
		& $0.95\sim1.00$ & Uniform \\

		$A_{\mathrm{mach}}$
		& $0.05\sim0.10$ & Uniform \\
		
		$H_{\mathrm{mach}}$
		& $8$ & Fixed \\
		
		$\alpha_{\mathrm{mach}}$
		& $1.10\sim1.15$ & Uniform \\
		
		$p$
		& $64$ & Fixed \\
		
		$f_b$
		& $55-65$ & Uniform \\
		
		$f_c$
		& $340-360$ & Uniform \\
		
		$\xi$
		& $0.95\sim1.00$ & Uniform \\
		
		$\eta$
		& $0.08\sim0.10$ & Uniform \\
		
		$|\Delta f_s|$
		& $\{0.25\sim1\}/60$
		& Uniform \\
		
		\bottomrule
	\end{tabularx}
\end{table}


Four known vessel types, namely cargo ships, passenger ships, fishing vessels, and motorboats, are considered in the simulation, together with synthetic outliers representing unknown-class samples. The vessel-specific identity parameters and shared source-generation settings are summarized in Tables~\ref{tab:vessel_parameters} and~\ref{tab:source_params}, respectively. Simulated vessel identities are instantiated by uniformly sampling the corresponding identity parameters from their specified ranges. For outliers, broader parameter ranges are used, and parameter combinations overlapping with the known classes are rejected and resampled. These outliers are included in training with the vessel type labeled as unknown, while retaining their physical-attribute annotations. The shared source-generation settings in Table~\ref{tab:source_params} introduce additional acoustic variability without acting as deterministic vessel-type cues.

\begin{figure*}[htbp]
	\centering
	\includegraphics[width=0.95\textwidth]{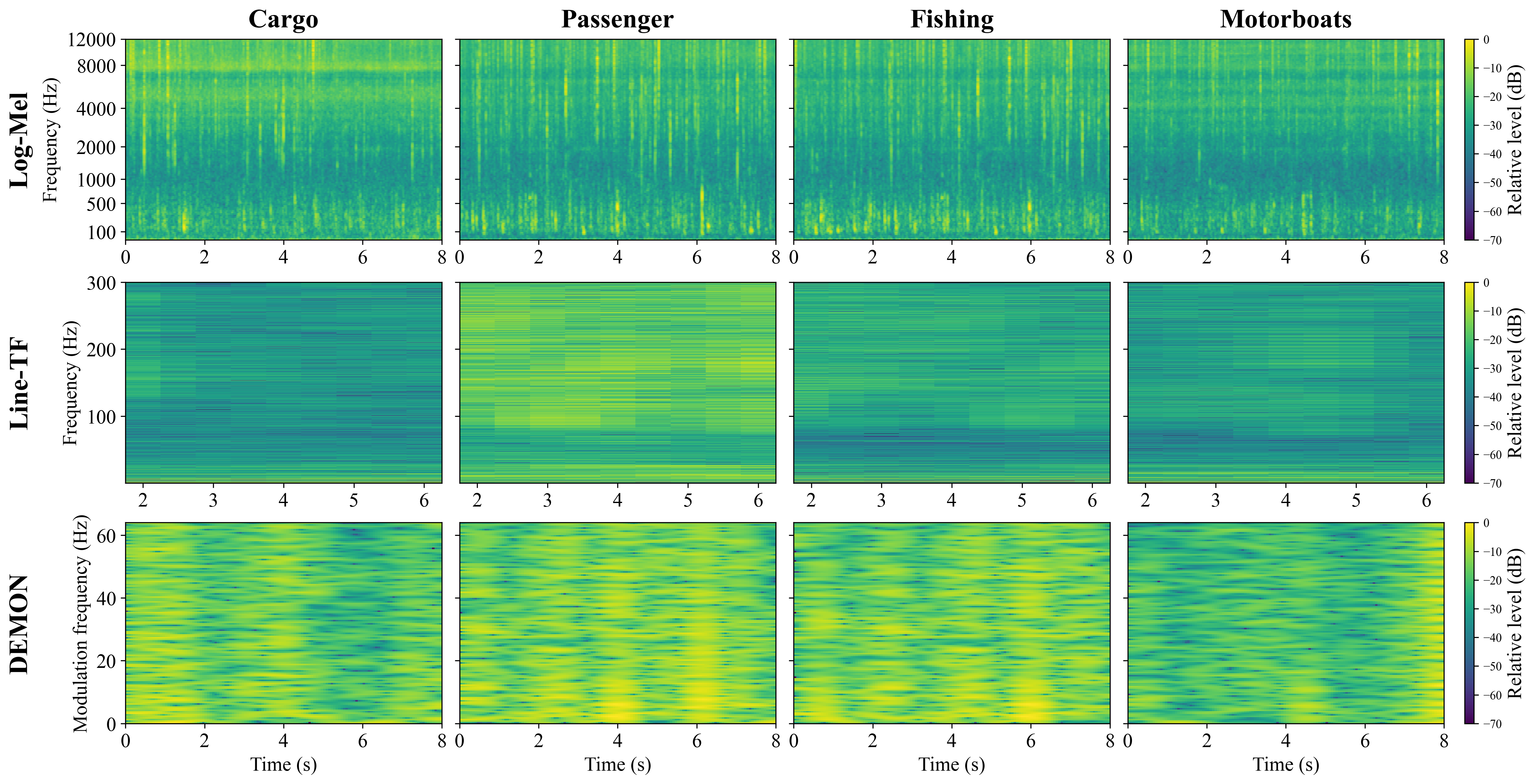}
	\caption{Examples of the Log-Mel, Line-TF, and DEMON representations for the four simulated vessel type under the accelerating operating condition.}
	\label{fig:4ships}
\end{figure*}

For each vessel type, including synthetic outliers, 400 simulated identities are generated, and four source signals are synthesized for each identity, corresponding to four distinct operating states. All source signals have a duration of 8~s and a sampling rate of 32~kHz. For each source signal, three independent realizations of the propagation channel and background noise are sampled within each of two SNR ranges, namely $[-5,5]$~dB and $[-20,-5]$~dB. This yields 24{,}000 recordings per SNR range and 48{,}000 recordings in total. The dataset is partitioned into training, validation, and test sets at a ratio of 8:1:1 using stratified group splitting based on vessel identity, ensuring that all recordings derived from a given vessel identity belong to the same subset and preventing identity-level information leakage. Representative acoustic features of the four simulated vessel classes are illustrated in Fig.~\ref{fig:4ships}.

\section{Experiments}
\label{sec:experiments}
\subsection{Experimental Setup}
\label{sec:setup}


\subsubsection{Implementation details}
For feature extraction, the Log-Mel spectrogram uses an FFT size of 4096 and a hop length of 320 with 128 Mel filters over 10--12{,}000~Hz. The Line-TF is computed at 1{,}024~Hz sampling with an FFT size of 4096 over 0.5--300~Hz. The DEMON spectrum is extracted from a 100--6{,}000~Hz band-passed envelope analyzed at 256~Hz over 0--64~Hz modulation frequency. Non-overlapping $16 \times 16$ patches are used for the Log-Mel and DEMON representations, while the Line-TF employs $16 \times 1$ frequency--time patches with a stride of $16 \times 2$, yielding 400, 380, and 64 tokens respectively. The encoder backbone is initialized with pretrained AudioMAE ViT-Base weights. The LLM is Qwen3-4B-Instruct, loaded in 4-bit precision and adapted with LoRA of rank 16.

\subsubsection{Training protocol}

Training was conducted in two stages. In the first stage, UPTE was optimized under Eq.~\ref{eq:uptloss} for 40 epochs using AdamW with a batch size of 16 and a weight decay of 0.05. The learning rates for the pretrained AudioMAE backbone and newly introduced modules were set to $2 \times 10^{-5}$ and $2 \times 10^{-4}$, respectively. The backbone was frozen for the first three epochs and subsequently unfrozen for joint optimization. The loss weights $\lambda_{\mathrm{soft}}$ , $\lambda_{\mathrm{peak}}$, $\lambda_{\mathrm{SFE}}$, $\lambda_{\mathrm{STE}}$, $\lambda_{\mathrm{BPFE}}$, and $\lambda_{\mathrm{MLD}}$ were set to 1.0, 0.5, 1.0, 1.0, 0.5, and 1.0, respectively.

In the second stage, the pretrained physics-aware encoder was integrated with the LLM for evidence-grounded description generation. The complete UPD framework was optimized under Eq.~\ref{eq:loss} for 3 epochs with an effective batch size of 16, while the pretrained LLM weights remained frozen and the LoRA adapters were trainable. The learning rates for the UPTE backbone, newly introduced UPTE modules, LLM projector, and LoRA adapters were set to $5 \times 10^{-6}$, $1 \times 10^{-5}$, $1 \times 10^{-4}$, and $2 \times 10^{-5}$, respectively, with a weight decay of 0.01. Both stages employed a cosine learning-rate schedule with a warmup ratio of 0.05. All experiments were conducted on an NVIDIA GeForce RTX 4090 GPU with 48~GB VRAM.

\subsection{Evaluation Protocol}
\label{sec:metrics}
\begin{figure*}[htbp]
	\centering
	
	\begin{subfigure}[t]{0.48\textwidth}
		\centering
		\includegraphics[width=\linewidth]{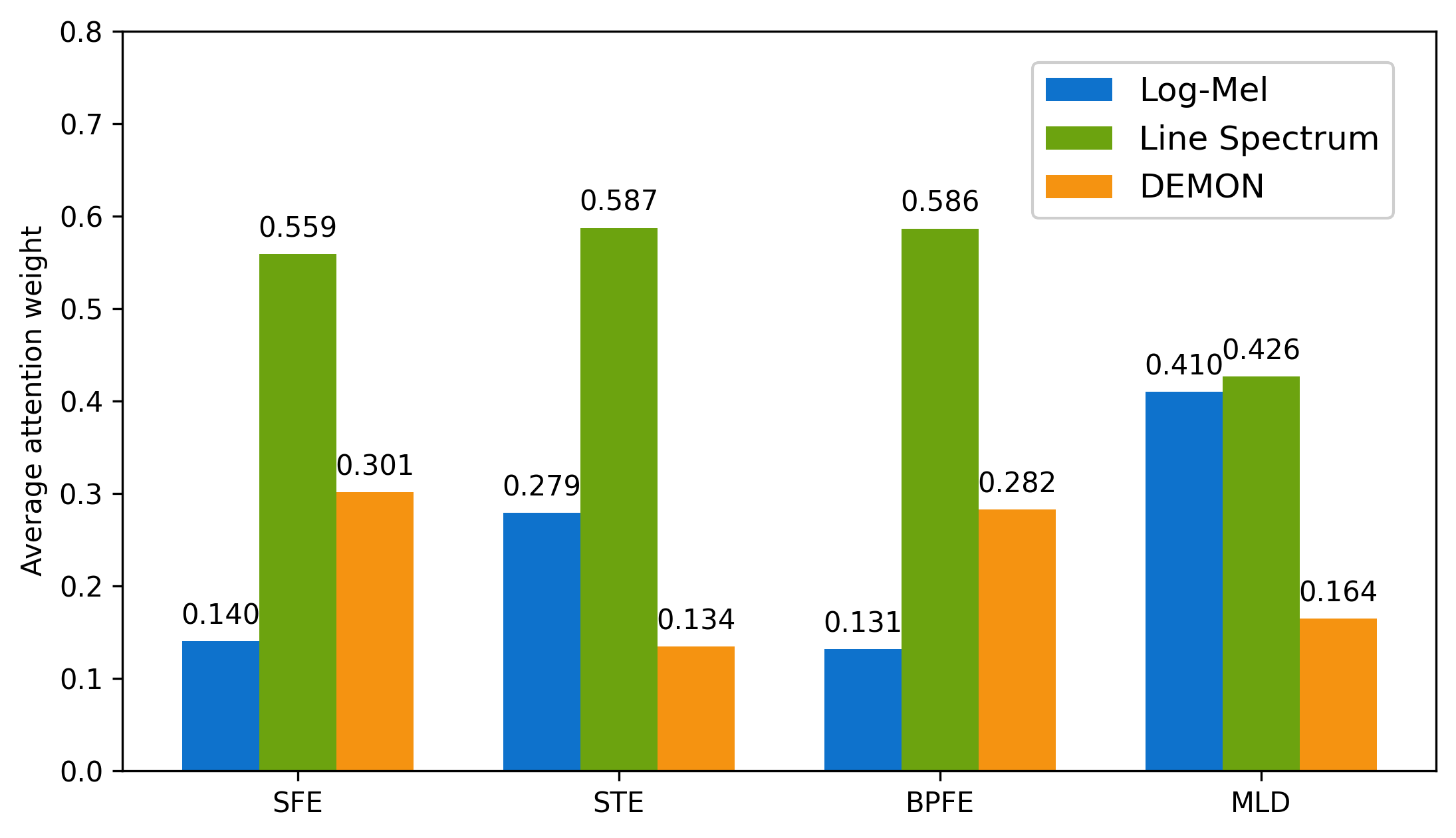}
		\caption{Average feature attention weights on the test set.}
		\label{fig:upt_attention}
	\end{subfigure}
	\hfill
	\begin{subfigure}[t]{0.48\textwidth}
		\centering
		\includegraphics[width=\linewidth]{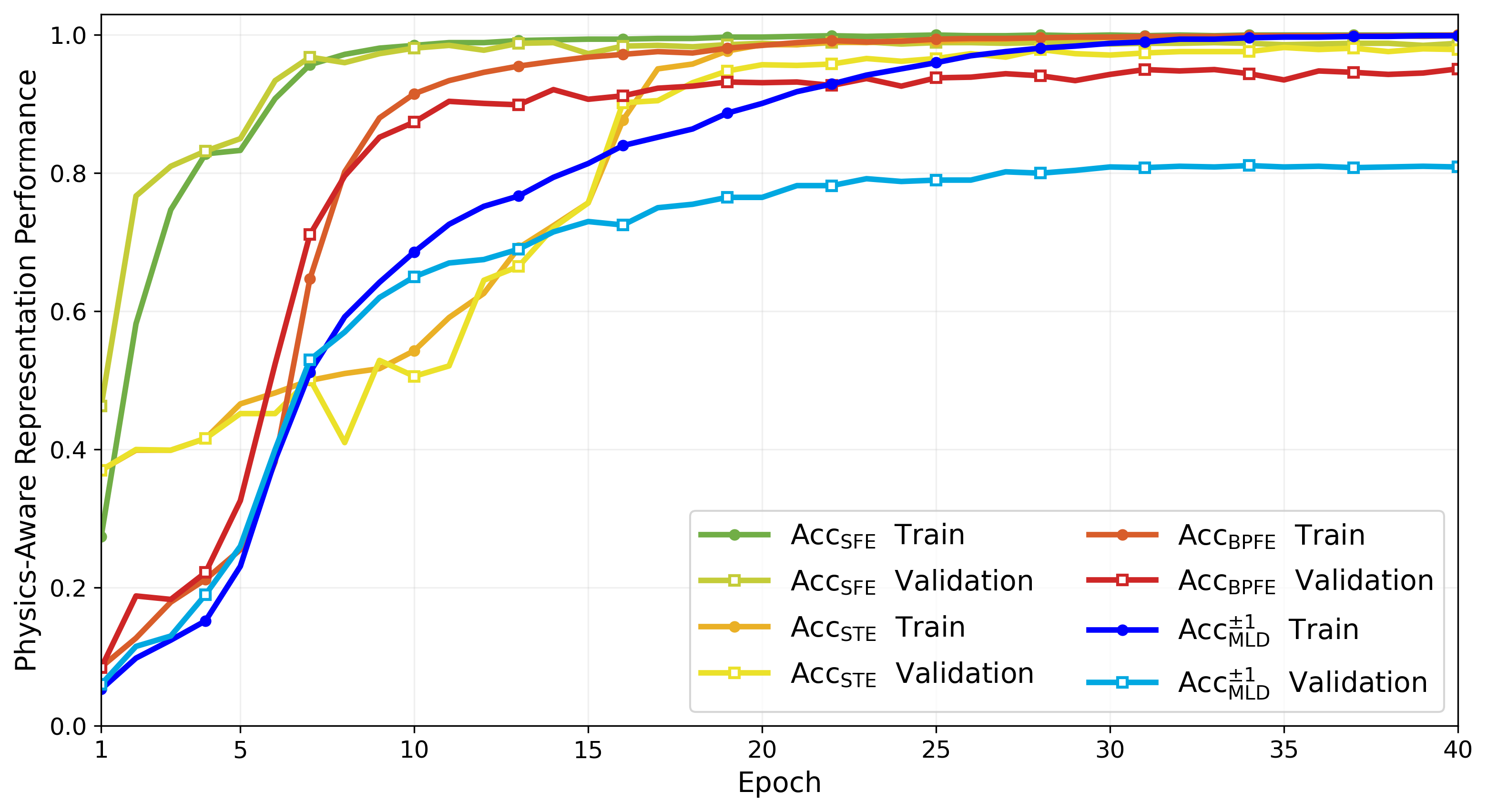}
		\caption{Training and validation performance on four physical tasks.}
		\label{fig:upt_task_performance}
	\end{subfigure}
	
	\caption{Analysis of the complete UPTE under the $-5$ to $5$ dB SNR condition.
		(a) Average attention weights assigned to the three acoustic
		representations for different physical tasks on the test set.
		(b) Evolution of training and validation performance for SFE,
		STE, BPFE, and MLD.}
	\label{fig:upt_analysis}
	
\end{figure*}

\begin{table*}[htbp]
	\centering
	\caption{Performance of UPTE under different input feature
		configurations and SNR conditions.}
	\label{tab:upt_ablation}
	\renewcommand{\arraystretch}{1.18}
	\setlength{\tabcolsep}{5.5pt}
	
	\begin{tabular}{>{\centering\arraybackslash}m{1.5cm}cccccccc}
		\toprule
		\multirow{2}{*}{\makecell{SNR range\\(dB)}}
		& \multicolumn{3}{c}{Input Features}
		& \multirow{2}{*}{\makecell{Task\\Attention}}
		& \multirow{2}{*}{$\mathrm{Acc}_{\mathrm{SFE}}$}
		& \multirow{2}{*}{$\mathrm{Acc}_{\mathrm{STE}}$}
		& \multirow{2}{*}{$\mathrm{Acc}_{\mathrm{BPFE}}$}
		& \multirow{2}{*}{$\mathrm{Acc}_{\mathrm{MLD}}^{\pm1}$} \\
		\cmidrule(lr){2-4}
		& Log-Mel & Line-TF & DEMON & & & & & \\
		\midrule
		
		\multirow{5}{*}{$-5$ to $5$}
		& \checkmark & -- & -- & --
		& 0.523 & 0.532 & 0.215 & 0.661 \\
		
		& -- & \checkmark & -- & --
		& 0.985 & 0.911 & 0.918 & 0.283 \\
		
		& -- & -- & \checkmark & --
		& 0.604 & 0.526 & 0.752 & 0.216 \\
		
		& \checkmark & \checkmark & \checkmark & --
		& \underline{0.986} & \underline{0.916}
		& \underline{0.931} & \underline{0.697} \\
		
		& \checkmark & \checkmark & \checkmark & \checkmark
		& \textbf{0.988} & \textbf{0.985}
		& \textbf{0.975} & \textbf{0.809} \\
		
		\midrule
		
		\multirow{5}{*}{$-20$ to $-5$}
		& \checkmark & -- & -- & --
		& 0.426 & 0.522 & 0.159 & 0.238 \\
		
		& -- & \checkmark & -- & --
		& 0.966 & 0.771 & 0.806 & 0.221 \\
		
		& -- & -- & \checkmark & --
		& 0.489 & 0.519 & 0.693 & 0.163 \\
		
		& \checkmark & \checkmark & \checkmark & --
		& \underline{0.974} & \underline{0.905}
		& \underline{0.823} & \underline{0.288} \\
		
		& \checkmark & \checkmark & \checkmark & \checkmark
		& \textbf{0.982} & \textbf{0.952}
		& \textbf{0.831} & \textbf{0.339} \\
		
		\bottomrule
	\end{tabular}
	
	\vspace{2pt}
	\parbox{\textwidth}{%
		\footnotesize
		\hspace{68pt}\textit{Note:} \textbf{Bold}: Best;
		\underline{Underlined}: Second best.
	}
	
\end{table*}

The evaluation follows the dual-granularity of the supervision. For the four physics-oriented tasks of UPTE, SFE, STE, and BPFE are evaluated using exact-match classification accuracy. For MLD, the two frequency bins with the highest predicted probabilities are selected and matched to the two ground-truth machinery frequencies in ascending order. Since small deviations between adjacent frequency bins may still represent physically accurate frequency localization, a tolerance-aware accuracy with a ±1 Hz threshold is adopted.

\begin{equation}
	\mathrm{Acc}_{q}
	=
	\frac{1}{N}
	\sum_{i=1}^{N}
	\mathbb{I}\left(
	\hat{y}_i^{(q)}=y_i^{(q)}
	\right),
	\quad
	q\in\{\mathrm{SFE},\mathrm{STE},\mathrm{BPFE}\}.
\end{equation}

\begin{equation}
	\mathrm{Acc}_{\mathrm{MLD}}^{\pm 1}
	=
	\frac{1}{2N}
	\sum_{i=1}^{N}
	\sum_{j=1}^{2}
	\mathbb{I}\left(
	\left|\hat{f}_{i,(j)}-f_{i,(j)}\right|
	\leq 1\,\mathrm{Hz}
	\right).
\end{equation}
where $N$ denotes the total number of test samples, and $\mathbb{I}(\cdot)$ is the indicator function, which equals 1 if the condition is satisfied and 0 otherwise. $y_i^{(q)}$ and $\hat{y}_i^{(q)}$ denote the ground-truth and predicted class labels of the $i^\text{th}$ sample for task $q$, respectively. $f_{i,(j)}$ and $\hat{f}_{i,(j)}$ represent the $j^\text{th}$ smallest ground-truth and predicted machinery fundamental frequencies of the $i^\text{th}$ sample, respectively, after sorting in ascending order.

For the structured descriptions of UPD, conventional sentence-level text similarity metrics do not reflect the correctness of individual physical attributes. We therefore adopt an \emph{attribute-level} protocol: each generated response is parsed into individual fields according to the fixed field order. Classification accuracy is used to evaluate vessel type $\mathrm{Acc}_{\mathrm{vessel}}$, operating state $\mathrm{Acc}_{\mathrm{state}}$, and shaft-frequency trend $\mathrm{Acc}_{\mathrm{trend}}$. For numerical attributes, including the initial, final shaft frequencies, the corresponding BPF values and SNR, mean absolute error (MAE) is calculated as
\begin{equation}
	\mathrm{MAE}=\frac{1}{N}\sum_{i=1}^{N}|\hat{y}_i-y_i|,
\end{equation}
where $\hat{y}_i$ and $y_i$ denote the predicted and ground-truth values of the $i^\text{th}$ sample, respectively. Lower MAE and RMSE indicate higher estimation accuracy.

For machinery fundamental frequencies, whose ordering carries no physical meaning, a sorting-based matching strategy is adopted. The predicted and reference frequency sets are independently sorted in ascending order, and the MAE is computed over the corresponding frequency pairs as
\begin{equation}
\mathrm{MAE}_{\mathrm{mach}}=
\frac{1}{2N}\sum_{i=1}^{N}\sum_{j=1}^{2}
\left|\hat f_{i,(j)}-f_{i,(j)}\right|,
\end{equation}
where $\hat f_{i,(j)}$ and $f_{i,(j)}$ denote the $j^\text{th}$ smallest predicted and ground-truth machinery frequencies of the $i^\text{th}$ sample, respectively. This provides a permutation-invariant measure of frequency estimation accuracy without imposing a frequency tolerance threshold.

\subsection{Physics-Oriented Task Performance}
\label{sec:upt_results}

\begin{table*}[htbp]
	\centering
	\caption{Comparison of UPD with different acoustic front-ends and
		adaptation strategies under different SNR conditions.}
	\label{tab:upd_comparison}
	\renewcommand{\arraystretch}{1.18}
	\setlength{\tabcolsep}{3.8pt}
	\small
	
	\begin{tabular}{ccccccccccc}
		\toprule
		\multirow{2}{*}{SNR (dB)} &
		\multirow{2}{*}{Method} &
		\multirow{2}{*}{$\mathrm{Acc}_{\mathrm{vessel}}$} &
		\multirow{2}{*}{$\mathrm{Acc}_{\mathrm{state}}$} &
		\multirow{2}{*}{$\mathrm{Acc}_{\mathrm{trend}}$} &
		\multicolumn{2}{c}{$\mathrm{MAE}_{f_s}$ (Hz)} &
		\multicolumn{2}{c}{$\mathrm{MAE}_{\mathrm{BPF}}$ (Hz)} &
		\multirow{2}{*}{\shortstack{$\mathrm{MAE}_{\mathrm{mach}}$\\(Hz)}} &
		\multirow{2}{*}{\shortstack{$\mathrm{MAE}_{\mathrm{SNR}}$\\(dB)}} \\
		\cmidrule(lr){6-7}
		\cmidrule(lr){8-9}
		& & & & & Initial & Final & Initial & Final & & \\
		\midrule
		
		\multirow{6}{*}{$-5$ to $5$}
		
		& Random
		& 0.249 & 0.263 & 0.331
		& 0.430 & 0.4453
		& 2.850 & 2.872
		& 48.127
		& 4.125 \\
		
		& Audio-Flamingo-Next (Zero-shot)
		& 0.250 & 0.250 & \underline{0.500}
		& 0.711 & 1.290
		& 4.959 & 11.542
		& 58.039
		& 2.433 \\
		
		& Audio-Flamingo-Next (LoRA)
		& 0.250 & 0.254 & \underline{0.500}
		& \underline{0.297} & \underline{0.300}
		& 2.019 & 2.031
		& 45.861
		& \underline{1.120} \\
		
		& Qwen2.5-Omni (Zero-shot)
		& 0.249 & 0.250 & \underline{0.500}
		& 0.717 & 0.812
		& 4.961 & 11.535
		& 78.748
		& 3.455 \\
		
		& Qwen2.5-Omni (LoRA)
		& \underline{0.282} & \underline{0.270} & \underline{0.500}
		& 0.310 & 0.309
		& \underline{1.691} & \underline{1.705}
		& \underline{41.607}
		& \textbf{0.693} \\
		
		& UPD with Complete UPTE
		& \textbf{0.974} & \textbf{0.877} & \textbf{0.976}
		& \textbf{0.014} & \textbf{0.015}
		& \textbf{0.146} & \textbf{0.151}
		& \textbf{1.533}
		& 2.420 \\
		
		\midrule
		
		\multirow{6}{*}{$-20$ to $-5$}
		
		& Random
		& 0.248 & 0.245 & 0.334
		& 0.429 & 0.442
		& 2.817 & 2.864
		& 52.893
		& 4.979 \\
		
		& Audio-Flamingo-Next (Zero-shot)
		& 0.250 & 0.250 & \underline{0.500}
		& 0.711 & 1.291
		& 4.959 & 11.542
		& 58.039
		& 4.082 \\
		
		& Audio-Flamingo-Next (LoRA)
		& 0.250 & \underline{0.251} & \underline{0.500}
		& 0.302 & \underline{0.305}
		& \underline{1.694} & \underline{1.710}
		& \underline{43.573}
		& \underline{1.956} \\
		
		& Qwen2.5-Omni (Zero-shot)
		& 0.252 & 0.248 & \underline{0.500}
		& 0.707 & 0.706
		& 4.956 & 11.534
		& 79.783
		& 4.172 \\
		
		& Qwen2.5-Omni (LoRA)
		& \underline{0.260} & 0.250 & \underline{0.500}
		& \underline{0.301} & 0.308
		& 1.789 & 2.240
		& 45.875
		& \textbf{0.859} \\
		
		& UPD with Complete UPTE
		& \textbf{0.861} & \textbf{0.741} & \textbf{0.891}
		& \textbf{0.017} & \textbf{0.018}
		& \textbf{0.729} & \textbf{0.728}
		& \textbf{11.295}
		& 2.742 \\
		
		\bottomrule
	\end{tabular}
	
	\vspace{2pt}
	\parbox{\textwidth}{%
		\footnotesize
		\textit{Note:} \textbf{Bold}: Best;
		\underline{Underlined}: Second best.
	}
	
\end{table*}

Table~\ref{tab:upt_ablation} summarizes the performance of UPTE under different input representations and SNR conditions. All three representations provide useful information for physical-attribute estimation, exhibiting distinct strengths across tasks. Under the $-5$ to $5$ dB SNR condition, Line-TF achieves the best single-representation performance on SFE, STE, and BPFE, with accuracies of 0.985, 0.911, and 0.918, respectively. This advantage stems from its high frequency resolution and ability to preserve narrowband tonal trajectories. In contrast, Log-Mel performs best on MLD, achieving an accuracy of 0.661, highlighting the importance of broader spectral context for MLD. While DEMON does not achieve the highest accuracy on any individual task, it outperforms Log-Mel on both SFE and BPFE, reflecting its ability to capture propeller-related periodic amplitude modulations that provide additional frequency cues. Similar trends are observed under the more challenging $-20$ to $-5$ dB condition, where Line-TF remains dominant on SFE, STE, and BPFE, Log-Mel achieves the highest MLD accuracy, and DEMON maintains relatively strong performance on SFE and BPFE. Notably, combining all three representations consistently outperforms any single representation across all four tasks under both SNR conditions, even without task-specific attention. This improvement is particularly evident for STE under low SNR, where accuracy increases from 0.771 with Line-TF alone to 0.905 with feature fusion. These results demonstrate that the three representations capture complementary acoustic characteristics, providing mutually beneficial information for physical-attribute estimation.

Introducing task-specific feature attention further improves performance across all four tasks under both SNR conditions. The average accuracy gain is more pronounced at higher SNR, with improvements of 0.2, 6.9, 4.4, and 11.2 percentage points for SFE, STE, BPFE, and MLD, respectively, compared with 0.8, 4.7, 0.8, and 5.1 percentage points at lower SNR. To examine how attention exploits the complementary representations, Fig.~\ref{fig:upt_attention} presents the average task-specific attention weights on the test set under the $-5$ to $5$ dB condition. The learned weights exhibit distinct task-dependent preferences that are consistent with the single-representation results in Table~\ref{tab:upt_ablation}. Line-TF receives the highest attention weights for SFE, STE, and BPFE, consistent with its superior individual performance on these tasks. Similarly, Log-Mel receives relatively high attention for MLD, while DEMON receives greater emphasis on SFE and BPFE. These results suggest that task-specific attention enables UPTE to adaptively emphasize the acoustic representations most relevant to each physical task.

Finally, Fig.~\ref{fig:upt_task_performance} presents the training and validation performance of the complete UPTE under the $-5$ to $5$ dB SNR condition. All four physics-oriented tasks show substantial improvements during the early training epochs. SFE, STE, and BPFE rapidly achieve high validation accuracies, demonstrating the effectiveness of UPTE in learning propulsion-related frequency characteristics. In contrast, MLD converges more gradually and reaches a lower accuracy, suggesting that identifying multiple machinery-related spectral components remains a more challenging task.

\subsection{Recognition and Evidence-Grounded Reasoning}
\label{sec:upd_results}

\begin{figure*}[htbp]
	\centering
	\includegraphics[width=0.95\textwidth]{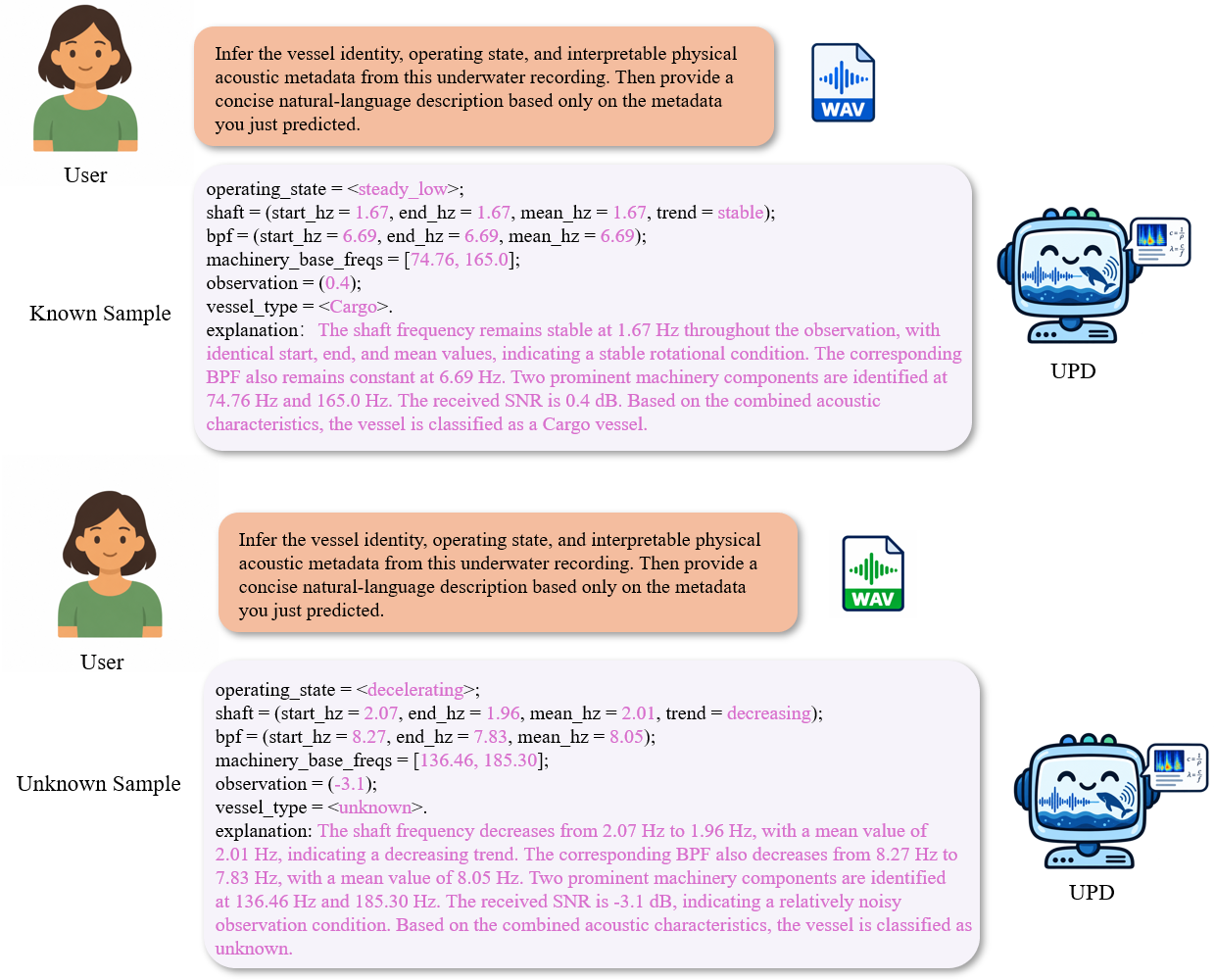}
	\caption{Illustrative input-output examples of the proposed UPD framework}
	\label{fig:UPDresults}
\end{figure*}

Table~\ref{tab:upd_comparison} compares UPD with a random baseline and two representative general-purpose LALMs, Qwen2.5-Omni and Audio-Flamingo-Next, each evaluated in both zero-shot and LoRA-adapted settings, under different SNR conditions. Unlike conventional prediction-based evaluation, all reported metrics in this subsection are extracted from the structured textual responses generated by the models, covering both physical-attribute descriptions and vessel recognition decisions. These two LALMs are selected for their strong classification performance in UA-Bench~\cite{sun2026uabench}. In the zero-shot setting, the original pretrained models are evaluated with the prompt explicitly specifying candidate categories for categorical attributes and valid ranges for frequency-related attributes. In the adapted setting, the models are fine-tuned on the SAD dataset by optimizing its projection module and LoRA parameters while keeping the pretrained audio encoder frozen.

The accuracy of generated physical evidence is first evaluated through physical-attribute estimation. Compared with the best-performing LoRA-adapted LALM, UPD reduces frequency-estimation MAEs by over 90\% under the $-5$ to $5$ dB SNR condition. Under the more challenging $-20$ to $-5$ dB condition, the reductions remain substantial, ranging from approximately 57\% to 94\%. These improvements demonstrate that physics-aware representations enable UPD to generate more accurate descriptions of fine-grained vessel-related acoustic characteristics, even under strong background noise. Notably, both LoRA-adapted LALMs achieve lower SNR estimation errors than UPD, suggesting that the advantages of physics-aware representations are more pronounced for vessel-specific acoustic attributes than for SNR estimation.

Building on the generated physical evidence, UPD also achieves substantially higher vessel recognition accuracy across both SNR conditions. Under the $-5$ to $5$ dB condition, UPD achieves accuracies of 0.974, 0.877, and 0.976 for vessel type, operating state, and shaft-frequency trend, respectively. In comparison, both zero-shot LALMs perform near the random baseline for vessel type and operating state, while their trend accuracy of 0.500 reflects consistent predictions of the \textit{stable} class. Even after domain adaptation, categorical recognition improves only marginally. Under the more challenging $-20$ to $-5$ dB condition, UPD maintains accuracies of 0.861, 0.741, and 0.891, respectively, demonstrating substantially greater robustness than the general-purpose model. Together with the accurate physical-attribute descriptions, these results support the effectiveness of the proposed feature--evidence--decision framework for underwater acoustic recognition.

Fig.~\ref{fig:UPDresults} further illustrates the evidence-grounded reasoning process through representative user--model interactions. Given an acoustic recording and a natural-language query, UPD first articulates the underlying physical evidence through structured acoustic attributes and then provides a corresponding recognition decision. The first example demonstrates cargo-ship recognition supported by propulsion- and machinery-related acoustic characteristics. The second example shows that, when the input does not belong to any of the four predefined vessel categories, UPD can still extract and describe its physical attributes while assigning the vessel type as \texttt{unknown}. These examples illustrate how UPD integrates physical evidence and recognition decisions within a unified, interpretable textual response.

\subsection{Evidence Intervention Analysis}
\begin{figure*}[htbp]
	\centering
	\includegraphics[width=0.95\textwidth]{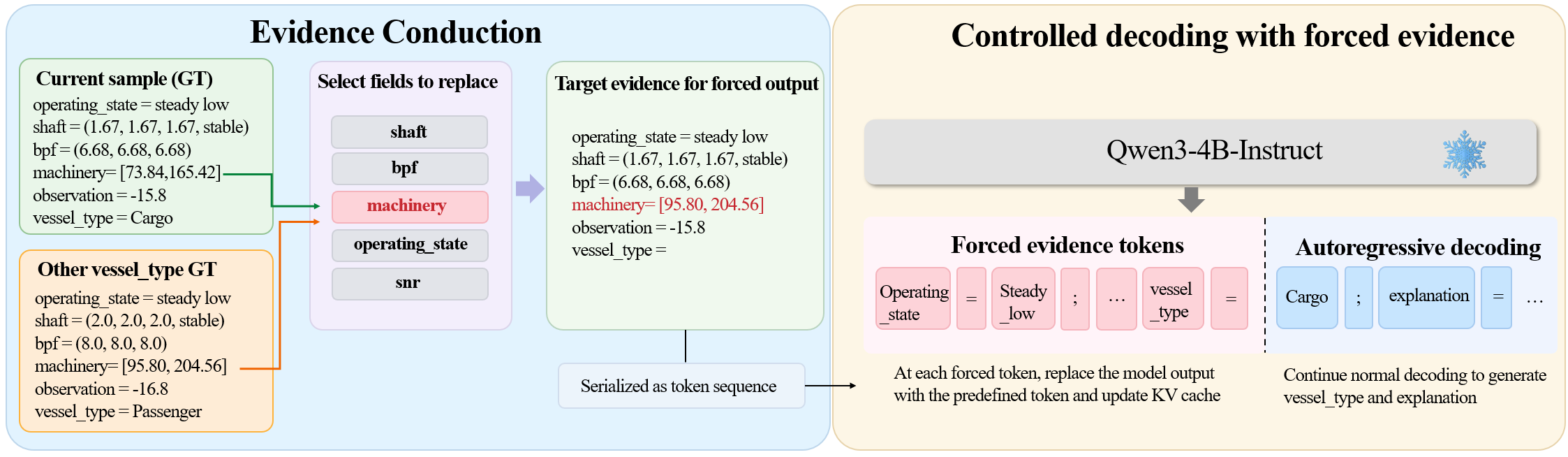}
	\caption{Illustration of the evidence intervention experiment. Selected physical evidence fields are replaced while the acoustic input remains unchanged to examine their influence on the final vessel recognition decision.}
	\label{fig:evidence_intervention_overall}
\end{figure*}

\begin{figure*}[t]
	\centering
	\begin{minipage}[t]{0.49\textwidth}
		\centering
		\includegraphics[width=\linewidth]{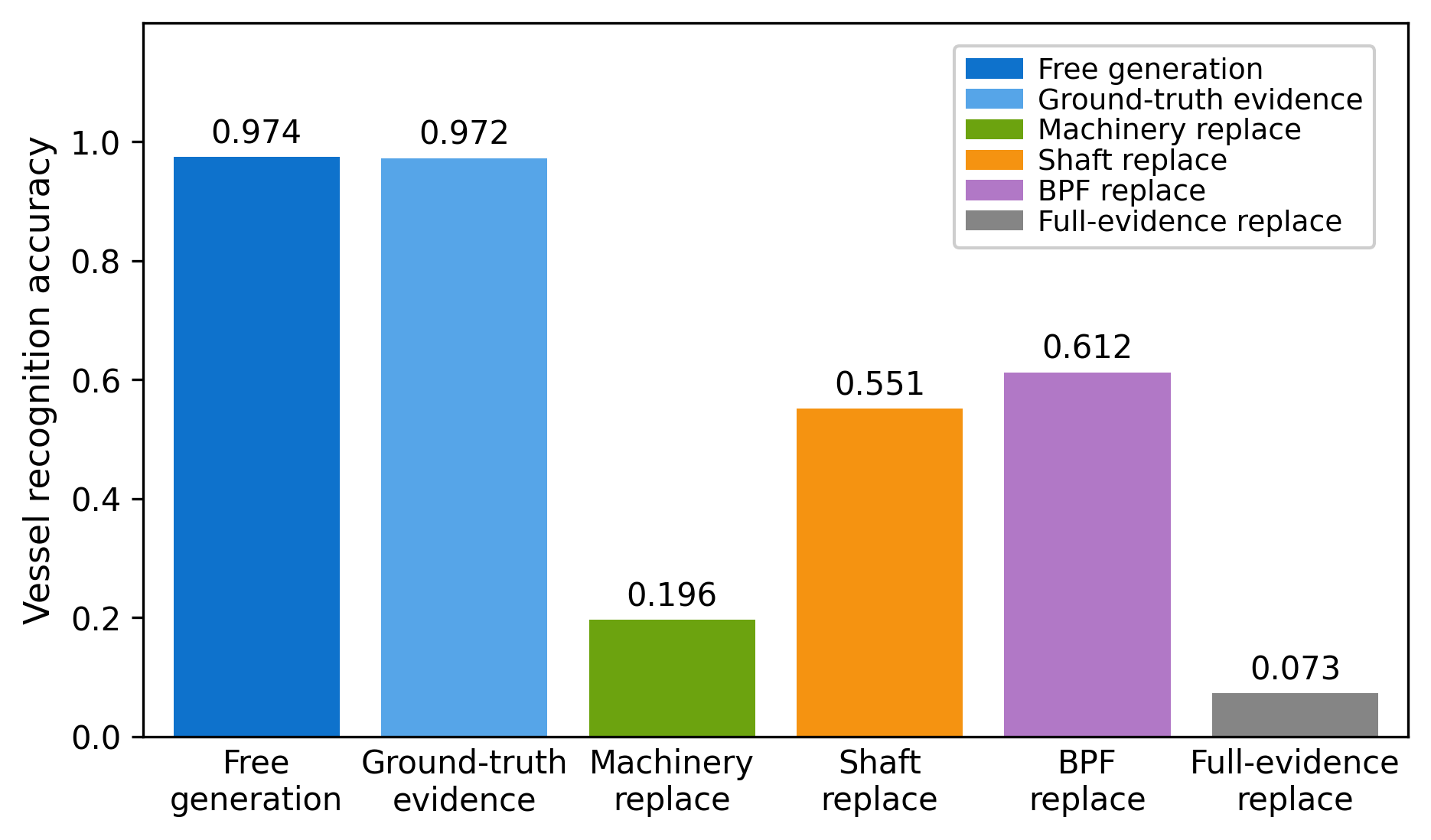}
		\par\smallskip
		(a) Evidence intervention under $-5$ to $5$ dB
	\end{minipage}\hfill
	\begin{minipage}[t]{0.49\textwidth}
		\centering
		\includegraphics[width=\linewidth]{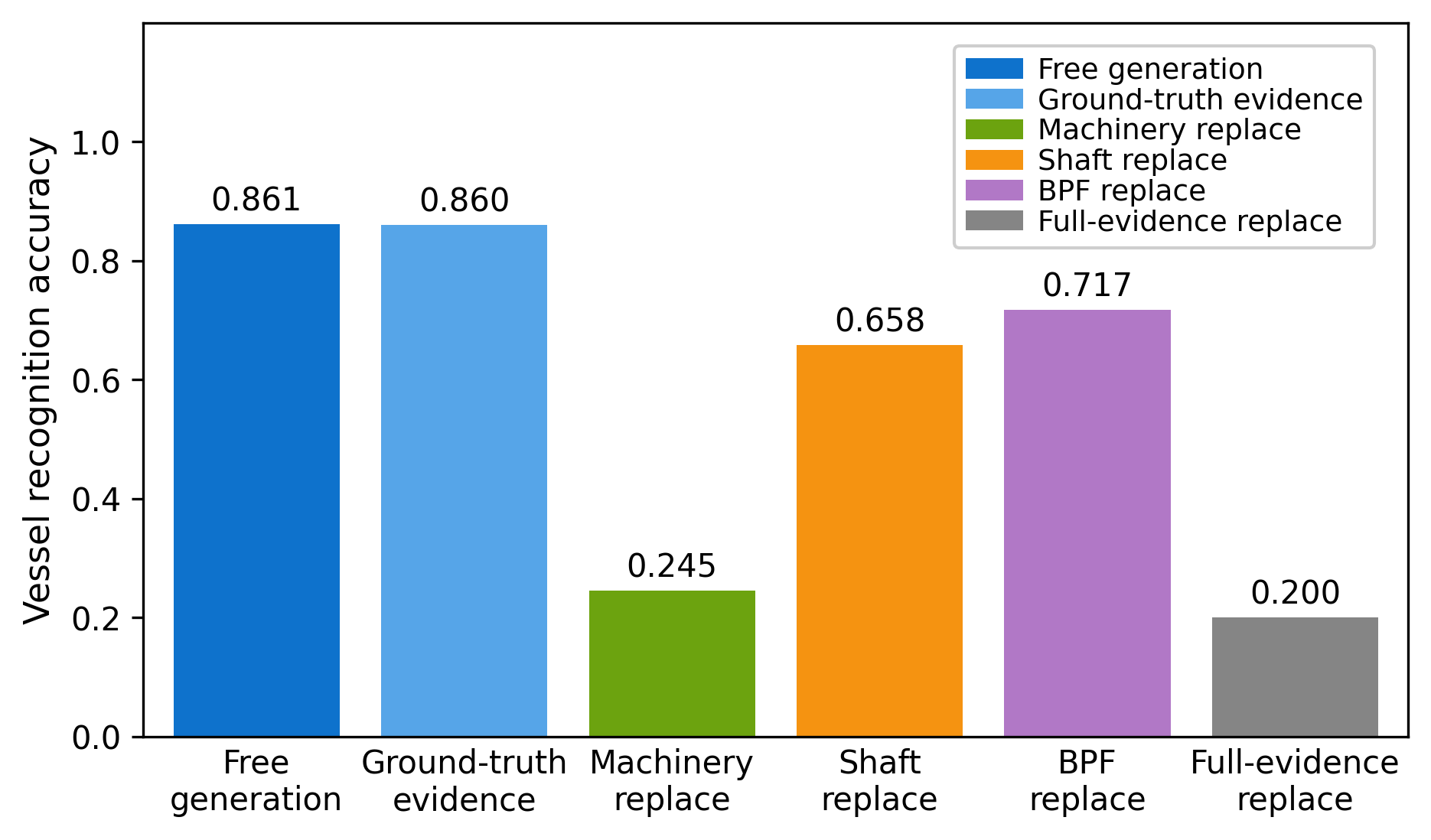}
		\par\smallskip
		(b) Evidence intervention under $-20$ to $-5$ dB
	\end{minipage}
	\caption{Effects of evidence interventions on vessel recognition accuracy under different SNR conditions. }
	\label{fig:evidence_intervention}
\end{figure*}

\begin{table*}[htbp]
	\centering
	\caption{Ablation study of token configurations and
		UPTE training settings under different SNR conditions.}
	\label{tab:architecture_ablation}
	
	\resizebox{\textwidth}{!}{%
		\begin{tabular}{
				c
				@{\hspace{5pt}}c
				@{\hspace{5pt}}c
				@{\hspace{7pt}}c
				c c c c c c c c c
			}
			\toprule
			\multirow{2}{*}{SNR (dB)}
			& \multirow{2}{*}{\shortstack{CLS \\tokens}}
			& \multirow{2}{*}{\shortstack{Audio \\tokens}}
			& \multirow{2}{*}{UPTE}
			& \multirow{2}{*}{$\mathrm{Acc}_{\mathrm{vessel}}$}
			& \multirow{2}{*}{$\mathrm{Acc}_{\mathrm{state}}$}
			& \multirow{2}{*}{$\mathrm{Acc}_{\mathrm{trend}}$}
			& \multicolumn{2}{c}{MAE$_{f_s}$ (Hz)}
			& \multicolumn{2}{c}{MAE$_{\mathrm{BPF}}$ (Hz)}
			& \multirow{2}{*}{\shortstack{MAE$_{\mathrm{mach}}$\\(Hz)}}
			& \multirow{2}{*}{\shortstack{MAE$_{\mathrm{SNR}}$\\(dB)}} \\
			\cmidrule(lr){8-9}
			\cmidrule(lr){10-11}
			& & & & & & &
			Initial & Final & Initial & Final & & \\
			\midrule
			
			\multirow{4}{*}{$-5$ to $5$}
			& \checkmark & \checkmark & Frozen
			& \underline{0.973} & 0.872 & \underline{0.970}
			& 0.021 & 0.023 & 0.249 & 0.258
			& 2.595 & 2.580 \\
			
			& \checkmark & -- & Jointly Trained
			& 0.970 & \underline{0.876} & \textbf{0.976}
			& \underline{0.015} & 0.023
			& \underline{0.176} & \underline{0.183}
			& \textbf{1.497} & 3.020 \\
			
			& -- & \checkmark & Jointly Trained
			& 0.969 & 0.852 & 0.954
			& \textbf{0.014} & \underline{0.017}
			& 0.205 & 0.217
			& 1.573 & \underline{2.490} \\
			
			& \checkmark & \checkmark & Jointly Trained
			& \textbf{0.974} & \textbf{0.877} & \textbf{0.976}
			& \textbf{0.014} & \textbf{0.015}
			& \textbf{0.146} & \textbf{0.151}
			& \underline{1.533} & \textbf{2.420} \\
			
			\midrule
			
			\multirow{4}{*}{$-20$ to $-5$}
			& \checkmark & \checkmark & Frozen
			& 0.843 & \underline{0.741} & 0.890
			& 0.018 & 0.020
			& \underline{0.729} & 0.729
			& \underline{11.295} & 2.776 \\
			
			& \checkmark & -- & Jointly Trained
			& \textbf{0.862} & \underline{0.741} & 0.889
			& \textbf{0.014} & \textbf{0.016}
			& \textbf{0.605} & \textbf{0.611}
			& \textbf{11.206} & 3.178 \\
			
			& -- & \checkmark & Jointly Trained
			& 0.859 & \textbf{0.794} & \textbf{0.895}
			& \underline{0.017} & \underline{0.018}
			& 1.223 & 1.227
			& 11.339 & \textbf{2.590} \\
			
			& \checkmark & \checkmark & Jointly Trained
			& \underline{0.861} & \underline{0.741}
			& \underline{0.891}
			& \underline{0.017} & \underline{0.018}
			& \underline{0.729} & \underline{0.728}
			& \underline{11.295} & \underline{2.742} \\
			
			\bottomrule
		\end{tabular}%
	}
	
	\vspace{2pt}
	\parbox{\textwidth}{%
		\footnotesize
		\textit{Note:} \textbf{Bold}: Best;
		\underline{Underlined}: Second best.
	}
	
\end{table*}

To investigate the influence of physical evidence on vessel recognition decisions, we conduct an evidence intervention experiment by selectively replacing generated evidence fields while keeping the acoustic input unchanged, as illustrated in Fig.~\ref{fig:evidence_intervention_overall}. Six settings are considered: Free generation, where the model generates its output without intervention; Ground-truth evidence, where the generated evidence is replaced with ground-truth attributes of the corresponding recording; Machinery, Shaft, and BPF replacement, where the respective physical evidence fields are individually replaced with deliberately incorrect attributes; and Full-evidence replacement, where all physical evidence fields are replaced simultaneously.

Fig.~\ref{fig:evidence_intervention} presents the $\mathrm{Acc}_{\mathrm{vessel}}$ under different evident intervention. Replacing the generated evidence with ground-truth attributes yields nearly identical performance to free generation. This suggests that the final decision is conditioned on both the generated evidence and the acoustic input tokens. Consequently, replacing the generated attributes with ground-truth values offers little additional benefit. In contrast, replacing individual evidence fields with deliberately incorrect values leads to substantial performance degradation, confirming that physical evidence actively contributes to vessel recognition rather than merely serving as an explanatory output. Among the individual interventions, machinery-related evidence has the greatest influence, highlighting its dominant discriminative role in vessel identification. Full-evidence replacement produces the largest degradation, indicating that the physical attributes provide complementary information and that their joint contribution is important to recognition. Notably, post-intervention accuracy is consistently higher under low-SNR conditions despite the lower baseline accuracy. This suggests that the contribution of explicit physical evidence may vary with signal quality. In particular, the model may rely differently on acoustic representations and physical attributes as their respective reliability changes with SNR. Overall, these findings support the role of intermediate physical evidence as an integral component of the proposed feature--evidence--decision chain.

\subsection{Ablation Study}

To investigate the effects of joint UPTE optimization and different input-token configurations, we conduct ablation experiments under two SNR conditions, as summarized in Table~\ref{tab:architecture_ablation}. Compared with frozen UPTE, the full model with joint UPTE optimization achieves higher vessel recognition accuracy, with a more pronounced improvement under the lower-SNR condition. This suggests that joint optimization facilitates the adaptation of acoustic representations to downstream recognition tasks, particularly in challenging acoustic conditions. We further examine the contributions of CLS tokens and acoustic tokens. Under the $-5$ to $5$ dB condition, combining both token groups achieves the highest vessel recognition accuracy and generally provides the best physical-attribute estimation performance, highlighting their complementary contributions. Under the $-20$ to $-5$ dB condition, the two token groups 	 exhibit different strengths across tasks, while their combination offers competitive overall performance. Taken together, these results demonstrate that CLS tokens and acoustic tokens capture complementary information, and their joint use provides a more balanced representation for vessel recognition and physical-attribute estimation across different SNR conditions.

\section{Discussion and Conclusion}
\label{sec:conclusion}

\subsection{From description to evidence}
A central question is whether generated descriptions provide genuine physical evidence rather than merely fluent rationalizations. UPD addresses this concern through structured evidence-before-decision outputs, traceable physical supervision, and task-specific feature attention. The generated evidence can be verified against known physical parameters, while attention weights indicate task-dependent emphasis on different acoustic representations. Together, these analyses support both the physical verifiability of the generated evidence and its relevance to recognition decisions, providing a foundation for evidence-grounded underwater acoustic recognition.

\subsection{Perspectives}
Several directions emerge from this work. First, although the evidence intervention experiment demonstrates the sensitivity of recognition decisions to intermediate physical evidence, further analysis is needed to clarify how individual acoustic structures and generated evidence jointly contribute to decision-making. Second, extending the framework to real-world underwater acoustic recordings is an important direction for future research. Evaluating UPD on real-world datasets such as ShipsEar and DeepShip would help assess its generalization capability under realistic acoustic conditions. Third, incorporating calibrated confidence estimates for individual physical attributes would enable downstream users, including human operators and cooperating sensing nodes, to assess and weigh acoustic evidence according to its reliability. Finally, the flexible input layer of UPD provides opportunities to incorporate richer physical priors, including sound speed profiles, bathymetric information, and operational context, as textual priors. This extension could support environment-aware reasoning and bring underwater acoustic recognition closer to the analytical capabilities of experienced human operators.

\subsection{Conclusion}
\label{sec:conclusion_main}

This paper presented the UPD framework, which integrates physics-aware acoustic encoding with a pretrained LLM to enable evidence-grounded underwater acoustic recognition. Built upon a unified feature--evidence--decision chain, its recognition instantiation employs the proposed UPTE to learn task-disentangled physical representations, supported by a physics-parameterized data engine and dual-granularity supervision. Experiments on simulated recordings of four known vessel types demonstrated accurate physical-attribute estimation and vessel recognition through structured textual outputs, while synthetic unknown-class samples further demonstrated the framework's ability to characterize targets beyond the predefined categories. As the first part of a two-part series, this work establishes the recognition framework, while the companion Part~II extends it to low-SNR target detection with multichannel processing and environmental prediction.

\bibliographystyle{ieeetr}
\bibliography{ref}

\end{document}